\documentclass{article}

\usepackage[english]{babel}

\usepackage[letterpaper,top=2cm,bottom=2cm,left=3cm,right=3cm,marginparwidth=1.75cm]{geometry}

\usepackage{amsmath}
\usepackage{amssymb}
\usepackage{amsthm}
\usepackage{graphicx}
\usepackage[colorlinks=true, allcolors=blue]{hyperref}
\usepackage{authblk}
\usepackage{tikz}
\usetikzlibrary{quantikz2}
\usepackage{xcolor}
\usepackage{listings}
\usepackage{subcaption}
\usepackage[title]{appendix}

\renewcommand{\H}{\mathcal{H}}
\newtheorem{definition}{Definition}
\newtheorem{proposition}{Proposition}
\newtheorem{example}{Example}
\DeclareMathOperator{\diag}{diag}

\newcommand{\fig}[1]{\hyperref[fig:#1]{FIG~\ref*{fig:#1}}}

\author[1]{Vanio Markov}
\author[2]{Vladimir Rastunkov}
\author[2]{Daniel Fry}

\affil[1]{Wells Fargo}
\affil[2]{IBM Quantum, IBM Research}

\begin{document}
\title{Quantum Time Series Similarity Measures and Quantum Temporal Kernels}
\maketitle

\begin{abstract}
This article presents a quantum computing approach to designing of similarity measures and kernels for classification of stochastic symbolic time series. In the area of machine learning, kernels are important components of various similarity-based classification, clustering, and regression algorithms. An effective strategy for devising problem-specific kernels is leveraging existing generative models of the example space.
In this study we assume that a quantum generative model, known as quantum hidden Markov model (QHMM), describes the underlying distributions of the examples. The sequence structure and probability are determined by transitions within model's density operator space. Consequently, the QHMM defines a mapping from the example space into the broader quantum space of density operators. Sequence similarity is evaluated using divergence measures such as trace and Bures distances between quantum states. 
We conducted extensive simulations to explore the relationship between the distribution of kernel-estimated similarity and the dimensionality of the QHMMs Hilbert space. As anticipated, a higher dimension of the Hilbert space corresponds to greater sequence distances and a more distinct separation of the examples. To empirically evaluate the performance of the kernels, we defined classification tasks based on a simplified generative model of directional price movement in the stock market. We implemented two widely-used kernel-based algorithms \textemdash \ support vector machines and k-nearest neighbors \textemdash \ using both classical and quantum kernels. Across all classification task scenarios, the quantum kernels consistently demonstrated superior performance compared to their classical counterparts.
\end{abstract}

\section{Introduction}\label{sec:intro}
Many machine learning algorithms for time series use various measures of similarity between examples to solve a wide range of problems, including classification, regression, density estimation, and clustering. The measures of similarity play a critical role in the k-NN algorithms, feature-based algorithms, and kernel methods. Various measures are intended to capture different types of similarity within time series data. One category of similarities is estimated directly within the original time series space. In cases where precise timing and synchronization of data points are essential, time series are compared synchronously, in lock-step, using an one-to-one alignment of corresponding data points. 
Some common lock-step time series similarity measures include Euclidean distance, linear time warping (LTW)\cite{LTW01}, Pearson correlation coefficient \cite{Pearson}, Spearman rank correlation \cite{Spearman}. The synchronous methods are not applicable for series with different lengths and cannot capture the local dependencies among adjacent positions within time series data. These difficulties can be resolved by applying non-linear mappings between the time series which allow comparison of one to many points. This approach provides a degree of flexibility or ``warping" when aligning two time series. Such flexibility makes it possible to find optimal alignment between time series that may exhibit variations in timing, speed, or phase. Examples of these elastic similarity measures include the longest common sub-sequence (LCSS) \cite{TSClustering}, dynamic time warping (DTW) \cite{NIPS2001_a869ccbc}, edit distance \cite{Edit}, move-split-merge (MSM) distance \cite{MSM}, piecewise linear approximation (PLA) distance \cite{PLA}, symbolic aggregate approximation (SAX) distance \cite{Lin03}, elastic ensemble distance (EED) \cite{lines2015time}. The synchronous and elastic similarities are estimated over the entire time series and are often referred to as ``global similarities". Alternatively, some approaches measure similarity with respect to local but representative patterns known as ``shapelets" \cite{ye2011time} within time series. For each time series, the similarity between the time series and each shapelet in a ``dictionary" is calculated, resulting in distance features. Distance features may also be defined by global similarity measures. Any similarity measure is associated with a ``dissimilarity" feature space, where each time series is represented by its distance vector to the other time series.

Another group of highly successful, similarity-based methods, known as \textit{kernel methods} \cite{scholkopf2001learning} utilize a function that estimates the similarity between examples by computing the dot product of their images in a higher-dimensional feature space. When the kernel functions satisfy Mercer's theorem \cite{mercer1909functions, Vapnik1998} they can calculate the dot product directly in the example space, without explicitly construction of feature vectors \cite{boser1992training}. The objective of this method is to ensure that the high-dimensional feature representations of the examples are linearly separable with respect to their class labels \cite{Vapnik1998}. Kernel methods enable the construction of linear discriminative algorithms for learning domains with non-linear decision boundaries. Examples of such algorithms include support vector machines \cite{boser1992training}, kernel logistic and ridge regression \cite{murphy2012machine}, and kernel discriminant analysis \cite{Pekalska09}, among others. Kernel methods are also used in unsupervised learning algorithms to capture non-linear structures in data. These include kernel principal component analysis \cite{scholkopf1998nonlinear, PANDEY2023126639, scholkopf1999input}, non-linear independent component analysis \cite{bach2002kernel,schell2023nonlinear}, kernel K-means clustering \cite{shawe2004kernel}, kernel spectral clustering \cite{langone2016kernel}, and kernel density estimation \cite{HARVEY20123}. 

Most general kernel functions, such as polynomial functions, Gaussian radial basis functions, sigmoid kernels, and hyperbolic tangent functions, rely on simple pairwise distances, such as the Euclidean distance, between data points in the example space. However, these kernels may not be suitable when the example space contains time series with varying lengths. One direct solution to this problem is to utilize elastic similarity measures as demonstrated in the case of \textit{DTW kernels} \cite{lei2007astudy, Badiane2022}. If the goal is to estimate significant local similarities the method of local alignment score \cite{smith1981identification} can be used to define the kernel. This approach has been combined in \cite{LA04} with a convolution kernel \cite{haussler1999} to derive a \textit{local alignment kernel}.
Finally, for time series classification tasks involving symbolic, noisy, or high-dimensional data, it is appropriate to use a kernel based on the symbolic aggregate approximation (SAX) similarity measure \cite{Lin03}. 

The selection of kernel type for a specific machine learning task depends on the nature of the task and the specific characteristics of the process under consideration. It is important to establish a systematic method to define the kernel class and parameters for any particular learning scenario. 

One approach to kernel design involves leveraging information from an existing, unsupervised, generative model of the examples domain. Such models define underlying distributions within the examples domain. The underlying distributions define feature spaces where the examples are represented. There are multiple ways for implementing of generative model-based kernels.
The \textit{P-kernels} \cite{shawe2004kernel, haussler1999convolution} estimate the similarity between two time series based on the joint probability of their inference state sequences, also known as inference paths. These kernels assume the existence of a probabilistic generative model, such as a hidden Markov model or Gaussian mixture model. For a time series, the model defines probabilities of all its inference paths \textemdash \ sequences of model states or mixture components. This distribution serves as the kernel's feature space. In this approach, each time series is transformed into a feature vector of probabilities over all possible inference paths. The similarity between two series is then computed by taking the dot product of their respective feature vectors, with each component weighted by the probability of the corresponding inference path. Because the feature vector contains probabilities for all possible inference paths, these kernels are also referred to as \textit{marginalisation kernels}. 

In another generative approach, known as \textit{probability product kernel} \cite{Jebara2004ProbabilityPK}, each time series is mapped to a distribution within a parameterized class of distributions. The probabilistic models for the individual time series are identified using maximum likelihood estimation. The probability product kernel is defined as standard inner product between densities. Since the vectors in the feature space are probability distributions, the distance between them can also be estimated using common divergence measures like the Kullback-Leibler divergence, Jensen-Shannon divergence, or other statistical distance measures \cite{NIPS2003_0abdc563, JMLR:v6:cuturi05a}. It is important to note, that these kernel functions usually do not satisfy the Mercer's theorem, which can lead to various issues in kernel-based learning algorithms.

One of the first methods for exploiting the knowledge captured by a probabilistic model to define a similarity metrics is to embed the time series in the gradient space of the generative model. This approach is known as \textit{Fisher kernel} \cite{jaakkola1999fisher}. The gradient vectors of a series log-likelihood with respect to a model parameters describe how that parameters contribute to the generative process. These parameters can encompass transition and emission probabilities in the case of hidden Markov models, expectations, and variances in the context of Gaussian mixtures, and more. The gradient vector is known as \textit{Fisher score} \cite{jaakkola1999exploiting}. If two series have similar gradient vectors, it indicates the model generates them in a similar manner. Consequently, we consider the series being close to each other in terms of their generative processes. The Fisher kernel is defined as the inner product of the gradient vectors of any two time series.

In this article, we investigate similarity measures and their associated kernels for symbolic time series sampled from an underlying stochastic process language. We assume the existence of a Markovian quantum generative model of the language \cite{markov2023implementation}, which defines two classes of probability distributions: one over the observed sequences and the other over their inference paths of quantum states represented by quantum density operators. The feature spaces of our kernels are the spaces of quantum density operators. We estimate similarity in these spaces using divergence measures such as trace and Bures distances.

The article's organization is as follows: \textbf{section \ref{section:Preliminaries}} provides a brief introduction to concepts, definitions, and features related to stochastic process languages and the corresponding generative models. We formalize classification problems for symbolic time series based on the concept of a classification map. Depending on the domain of the classification map, we define two types of classification tasks: structural and predictive. We discuss the definitions and requirements of the kernel functions utilized in classification problems. In \textbf{section \ref{subsubsection:quantum_hmms}} we define a quantum generative model \textemdash \ quantum hidden Markov model (QHMM) \textemdash \ which serves as the foundation of our study. The model is defined as a completely positive trace-preserving (CPTP) map, enabling comprehensive definitions of symbolic sequence probabilities in the example space and generative state sequences in quantum density space. Subsequently, we discuss an unitary physical implementation of the model introduced in \cite{markov2023implementation}. In \textbf{section \ref{section:quantum_kernwls}}, we formally specify quantum predictive and structural kernels based on the introduced QHMM. We explain the intuition behind these kernels and establish some important probabilistic properties.
\textbf{Section \ref{section:empirical_evaluation}} presents the conducted simulations and empirical investigations regarding the impact of the QHMM’s Hilbert space dimension and the distance distributions in the context of classification tasks. Here, we empirically compare the performance of the introduced quantum kernels with their classical counterparts. The discussion of the physical implementation of the proposed kernels in the quantum circuits computing model is presented in \textbf{section \ref{section:quantum_kernels_circuits}}. The contributions of the article, its potential impact on the field, and the future research directions are presented in \textbf{section \ref{section:conclusion}}.

\section{\label{section:Preliminaries}Preliminaries}

\subsection{\label{subsubsection:stochastic_and_hmms}Stochastic Process Languages }

We consider a class of observable, discrete-time stationary stochastic processes denoted by
\begin{equation}
\label{y_t}
    \{ y_t : t \in \mathbb{N}, y_t \in\Sigma \},
\end{equation}
where $\Sigma=\{a_1, \ldots, a_m\}$ is a finite set of symbols, called an \textit{alphabet}. 

The set of all finite sequences over the alphabet $\Sigma$, including the empty sequence $\epsilon$, is denoted by ${\Sigma}^{*}$. The set of all sequences with length exactly $t$ is denoted by ${\Sigma}^{t}$. Any subset of ${\Sigma}^{*}$ is a \textit{language} ${L}$ over the alphabet. The sequences belonging to a language are referred to as \textit{words}. 

The set of sequences resulting from observations or measurements of the evolution of a discrete-time process is called a \textit{process language}. In this context, the index $t$ is non-negative and is interpreted as \textit{time} with the process sequences being considered as \textit{time series}.
It is straightforward to verify that if a word results from the observation of a process, then every one of its subwords has also been observed. Therefore, the process languages are \textit{subword-closed}.

A \textit{stochastic} process language $L$ is defined as a process language along with a set 
\begin{equation}
\label{eqn:stochstic language distribution_0}
 D^L=\bigl \{D_t^L : t \geq 0 \bigr \} 
\end{equation}
\noindent
of finite dimensional probability 
distributions $D_t^L$, each of which is defined on the sequences with length exactly $t$:

\begin{equation}
\label{eqn:stochstic language distribution}
    D_t^{L} = \bigl \{ P[\textbf{y}] : \textbf{y} \in {\Sigma}^{t}, \sum_{\textbf{y} \in {\Sigma}^{t} } P[\textbf{y}] = 1  \bigr \}
\end{equation}

\subsection{\label{subsubsection:probabilistic generative models} Probabilistic Generative Models}
We will consider probabilistic generative models for the stochastic process languages. These models describe the underlying distribution of the language assigning probability to each sequence from $\Sigma^*$. Examples of such models are 
variational autoencoders (VAEs), 
Gaussian mixture models (GMMs),
hidden Markov models (HMMs),
latent Dirichlet allocation (LDA),
Bayesian networks,
Boltzmann machines, etc \cite{GM2020100285}. 

A generative model of stochastic language can be defined assuming that each observation $y_t$ depends on the state of a non-observable or \textit{hidden} finite-state process 
 \begin{equation}
 \label{X_t}
 \{ x_t : t \in \mathbb{N}, x_t \in S \}, 
 \end{equation}
where $S =\{s_1 \ldots s_n\}$ are the process states.

\noindent
The joint process 
\begin{equation}
\label{y_t,x_t}
\{ y_t, x_t : t \in \mathbb{N},\,y_t \in {\Sigma},\, x_t \in S \}
\end{equation}
 is assumed to be stationary and is described by a linear model 
\begin{equation} 
\label{linear model}
\begin{array} {lcl}
    x_{t+1} = Ax_t\\
    y_{t} = Bx_t,
\end{array}
\end{equation}
where $A$ is a row-stochastic transition matrix and $B$ is column-stochastic emission matrix. 

At any moment in time $t$, the model is in a superposition (a stochastic mixture) of its hidden states, described by a stochastic vector $x_t \in \mathbb{R}^n$. The state component $x_t^i$ represents the probability of the 
process being in state $x_t=s_i$. The distribution of observable symbols at time $t$ is defined by the stochastic vector $y_{t} \in \mathbb{R}^m$ as the probability to observe $y_t=a_i$ is denoted by $y_t^i$. This model is known as a (classical) hidden Markov model (HMM).

For every HMM $\mathbf{M}$ we can define a set of \textit{observable operators} \textbf{T} as follows
\begin{equation}
\label{classic observable operators}
    \textbf{T} = \{ T_a : T_a=AB_a, a \in\Sigma \},
\end{equation}
where $B_a = \diag(B[a,i], i \in [1.n])$ are diagonal matrices and $B[a,i]$ represents the emission probability of observing symbol $a$ in state $s_i$ ~\cite{jaeger2005learning,carlyle_paz1971}.

Every component $T_{a}[i,j]$ of an observable operator $T_a$ defines the conditional probability to observe symbol $a$ when the process state evolves from state $s_i$ to $s_j$.

\begin{equation*}
    T_{a}[i,j] = P[s_j \mid s_i, a].
\end{equation*}
For any sequence $\textbf{a}={a_1} \ldots {a_t}$ we also can define an observable operator as follows: 
\begin{equation*}
    T_\textbf{a} = T_{a_l} \ldots T_{a_1}.
\end{equation*}
A HMM $\mathbf{M}$ defines probability of every observable sequence $\mathbf{a} = a_1 \ldots a_t$ as:
\begin{equation}
 \label{classic_observable_prob}
    P[\textbf{a} \vert \mathbf{M}]=\textbf{1}T_{a_t} \ldots T_{a_1}\textbf{x}_0=\textbf{1}T_{\textbf{a}}\textbf{x}_0,
\end{equation}
where \textbf{1} is the unit row vector with dimension $n$, and $\textbf{x}_0$ is the \textit{initial} state distribution of the model.
\noindent
With every model $\mathbf{M}$ we associate a \textit{sequence function} $f^{\mathbf{M}} : {\Sigma}^{*} \rightarrow \left[ 0, 1 \right] $ defined as:
\begin{equation}
    \label{eqn:hmm_sequence_function}
    {f^{\mathbf{M}}}(\textbf{a})=P(\textbf{a} \vert \mathbf{M}),\forall \textbf{a} \in {\Sigma}^{*}.
\end{equation}
\noindent
Through the function $f^{\mathbf{M}}$ the model defines a stochastic process language $L_M$ (\ref{eqn:stochstic language distribution}):

\begin{equation}
\label{eqn: D_M^T classic}
    D^{\mathbf{M}}_t = \{ {f^\mathbf{M}}(\textbf{a}) : \textbf{a} \in {\Sigma}^{t} \}
\end{equation}
\subsection{\label{subsubsection:classification problem} Classification Problem for Stochastic Languages}

We consider the \textit{symbolic time series classification problem} for a stochastic language $L$ and a finite set of class labels $C = \{ c_1, \ldots, c_k \}$. It is assumed that there exists an unknown \textit{probabilistic classification map} $\mathcal{P}: L \rightarrow P[C]$, which assigns to each sequence $\mathbf{y} \in L$ a probability distribution over the class labels.

We distinguish between two types of probabilistic classification maps:

\paragraph{Structural classification maps.}  
These maps, which we call \textit{structural}, assign class probabilities based solely on the structure of the observed sequence:
\begin{equation}
\label{clsmap_s}
p_s(\mathbf{y}, c) = P[c \mid \mathbf{y}], \quad \mathbf{y} \in L, \; c \in C.
\end{equation}
In structural classification, the class of a sequence depends only on its observed pattern. Examples include protein family classification based on amino acid sequences \cite{Dhanuka23}, detection of anomalous financial transactions \cite{Jurgovsky2018SequenceCF}, and intrusion detection in cybersecurity \cite{Odaini23}.

\paragraph{Predictive classification maps.}  
The second type of classification maps assign class probabilities depending on the expected \textit{future evolution} of the observed sequence. For a fixed horizon $k > 0$, the \textit{predictive classification map} is defined as:
\begin{equation}
\label{clsmap_p}
p_p(\mathbf{y}, c) = \sum_{\mathbf{z} \in \Sigma^k} P[\mathbf{z} \mid \mathbf{y}] \cdot P[c \mid \mathbf{yz}], \quad \mathbf{yz} \in L, \; c \in C,
\end{equation}
where $P[\mathbf{z} \mid \mathbf{y}]$ denotes the probability of observing continuation $\mathbf{z}$ of length $k$ after observing $\mathbf{y}$, and $P[c \mid \mathbf{yz}]$ is the class probability conditioned on $\mathbf{yz}$.

We refer to classification problems defined by (\ref{clsmap_p}) as \textit{predictive}. These problems arise in domains where future dynamics influence class membership, such as weather forecasting \cite{SharmaWeather21}, stock market trend prediction \cite{LiangSM23}, patient health monitoring \cite{Rosafalco20}, energy load forecasting \cite{Tsalikidis24}, and traffic flow analysis \cite{GOMES2023200268}.

\paragraph{Training and learning.}
A training dataset consists of labeled sequence samples:
\[
S = \left\{ (\mathbf{y}_i, c_i) \mid \mathbf{y}_i \in Y, \; c_i \in C, \; i \in [1, \ell] \right\},
\]
where $Y = \left\{ \mathbf{y}_i \sim D^L_i \right\}$ is a set of finite sequences sampled from respective distributions $D^L_i$ over $L$.
\noindent
The goal of the classification problem is to learn an approximation $\widehat{\mathcal{P}}$ to the true classification map $\mathcal{P}$.

\paragraph{Evaluation.}
The quality of the learned classifier $\hat{p}(\mathbf{y}, c)$ is typically evaluated using probabilistic scoring rules such as the Brier score, negative log-likelihood, or proper scoring metrics that assess the predictive distribution over labels. These measures quantify the fidelity of the learned stochastic classification behavior.

\subsection{\label{subsubsection:kernel functions} Kernel Functions }
A common approach to solving classification problems is to establish a similarity measure for pairs of examples that is consistent with their respective classes: examples within the same class should be considered closer according to this measure compared to examples from different classes. The similarity measure can be formalized as a \textit{kernel function}, which computes the similarity or distance between pairs of examples in the input domain. 
We will assume that the example domain is a stochastic process language $L$ as discussed in Section \ref{subsubsection:stochastic_and_hmms}. A kernel function is formally defined as follows:
\begin{equation}
\label{kernelfn}
K : L \times L \rightarrow \mathbb{R}, 
\end{equation}
where $K$ is a symmetric, positive semi-definite function.

\noindent
The idea behind this definition is to map the examples into a high-dimensional feature space where the data becomes more separable or even linearly separable, and then to compute similarity within that space. Let's consider an implicit or explicit map of the input examples to a high-dimensional Hilbert space $H$:
\begin{equation}
\label{map_fi}
\phi: L \rightarrow H.    
\end{equation}
\noindent\
If $\kappa$ is a symmetric positive semi-definite measure in the Hilbert space:

\begin{equation}
\label{kappa}
\kappa : H \times H \rightarrow \mathbb{R}, 
\end{equation}
\noindent
then the kernel function is specified as:
\begin{equation}
\label{kernel function}
K(x_1,x_2) = \kappa(\phi(x_1), \phi(x_2)) 
\end{equation}
\noindent

We construct a kernel function using the state-space of the stochastic generative model (\ref{linear model}). The model assigns probability to every observable sequence by equation (\ref{classic_observable_prob}). The model also defines a mapping between the examples $\mathbf{y} = y_1 \ldots y_t$ and the vectors of probabilities of observing an example $\mathbf{y}$ from each of the n possible states or inference paths:
\begin{equation}
 \label{classic_seq_map}
    \phi_{\mathbf{M}}(\textbf{y} )= T_{y_t} \ldots T_{y_1}\textbf{x}_0=T_{\textbf{y}}\textbf{x}_0
\end{equation}

construction of classification and regression learning algorithms that are nonlinear in the input space $L$ while having linear counterparts within the Hilbert space $H$.

In the next session we assume the existence of a quantum generative model for the language $L$ and apply this approach in the Hilbert space of the model.

\section{\label{subsubsection:quantum_hmms} Quantum Hidden Markov Models}

We will consider stochastic process languages defined by quantized HMMs known as quantum hidden Markov models (QHMMs) \cite{monras2011hidden}. A QHMM is a complete positive quantum operation \cite{nielsen_chuang_2010} or quantum channel, describing how a composite quantum system evolves according to its internal dynamics and simultaneously parts of it are observed by measurement. The model combines unitary hidden states evolution with the emission of observations correlated with the hidden states. The QHMM is a \textit{quantum stochastic generator}, formally defined as follows:

\begin{definition}[Quantum hidden Markov model~\cite{monras2011hidden}]
    \label{def_qhmm}
    A \textit{quantum} HMM (QHMM) $\mathbf{Q}$ over an $N$-dimensional Hilbert space $\H$ is a 4-tuple:
    \begin{equation}
    \label{qhmm} 
        \mathbf{Q} = \{ \Sigma,\, \H,\, \mathcal{T}=\{ T_a \}_{a \in \Sigma},\, \rho_0 \},
    \end{equation}
    where
    \begin{itemize}
    \item $\Sigma$ is a finite alphabet of observable symbols. 
    \item $\H$ is an $N$-dimensional Hilbert space. It defines the state-space of the model $\mathbf{Q} $ as the set of associated density operators $D(\H)$.
    
    \item $\mathcal{T}$ is a CPTP map (quantum channel) $\mathcal{T}: D(\H) \rightarrow D(\H)$. 
    \item $\{ T_a \}_{a \in \Sigma}, \sum_{a \in \Sigma}T_a^{\dagger}T_a=I_N$ is an operator-sum representation of $\mathcal{T}$ in terms of a complete set of Kraus operators $T_a$.
    \item $\rho_0$ is an initial state, $\rho_0 \in D(\H) $, where $D(\H)$ is the space of density operators defined in $\H$.
    \end{itemize}
\end{definition}
\noindent
 The space $D(\H)$ is a convex set, representing statistical ensembles of quantum states, which consists of all positive semi-definite operators $\rho$ on $\H$ with $\operatorname{tr}(\rho)=1$. When the map $\mathcal{T}=\{ T_a \}_{a \in \Sigma}$ is applied to a state $\rho \in D(\H)$, the model $\mathbf{Q}$ defines measurement outcome $a \in \Sigma$ with probability:
\begin{equation}
\label{sym_prob}
    P \bigl[ a  \vert \rho \bigr] = \operatorname{tr}(T_a \rho)
\end{equation}
\noindent
and the system's state after the measurement becomes:
\begin{equation}
\label{new_state}
    \rho_a = \frac{T_a \rho}{P\bigl[ a \vert \rho \bigr]}.
\end{equation}
The completeness of the set of Kraus operators $\{ T_a \}_{a \in \Sigma}$ guarantees that the measurement probabilities in each state define a distribution:

\begin{equation}
\label{eqn: D_Q^ro quantum}
    D^{\mathbf{Q}}_{\rho} = \{ P \bigl[ a  \vert \rho \bigr] : a \in \Sigma  \}.
\end{equation}

If the operation $\mathcal{T}$ is applied $t$ times starting at the initial state a sequence $\mathbf{y} = y_1 \ldots y_t$ will be observed with probability
\begin{equation}
    \label{eqn:seq_probability}
    P\bigl[ \mathbf{y} \vert \rho_0 \bigr] = \operatorname{tr}(T_{\mathbf{y}} \rho_0).
\end{equation}
\noindent
and the final state will be
\begin{equation}
\label{seq defined state}
    \rho_{\mathbf{y}} = \frac{T_{\mathbf{y}} \rho_{0}}{P\bigl[ \mathbf{y} \vert \rho_{0} \bigr]},
\end{equation}
\noindent
where
$$T_{\mathbf{y}}=T_{y_t} \ldots T_{y_1}$$
\noindent

For each QHMM $\mathbf{Q}$ the equation (\ref{eqn:seq_probability}) defines a \textit{sequence function} $f^{\mathbf{Q}}$ as follows:
\begin{equation}
    \label{eqn:qhmm_sequence_function}
    f^{\mathbf{Q}} (\mathbf{y}) = P\bigl[ \mathbf{y} \vert \rho_0 \bigr], \forall \mathbf{y} \in
    \Sigma^{*}. 
\end{equation}

The sequence function (\ref{eqn:qhmm_sequence_function}) defines a distribution $D^{\mathbf{Q}}_t$ over the sequences of length $t$ for $\forall t>0$, since the quantum operation $\mathcal{T}$ is completely positive:
\begin{equation}
\label{eqn: D_M^T quantum}
    D^{\mathbf{Q}}_t =  \bigl\{ {f^\mathbf{Q}}(\textbf{y}) : \textbf{y} \in {\Sigma}^{t}  \bigr\}
\end{equation}
\noindent
Therefore every QHMM $\mathbf{Q}$ defines a \textit{stochastic process language} $L^\mathbf{Q}$ (\ref{eqn:stochstic language distribution_0}) over the set of finite sequences $\Sigma^{*}$.

\begin{figure}[ht]

    \centering
    \begin{quantikz}[slice style=blue,wire types={q,q,q,q,q,n,c},row sep={20pt,between origins}]
        \lstick[3]{$\ket{s_0}$} & \gate[5][50]{U} &                                 &                &                                        &\gate[5][50]{U}    &                                 &        \ \ldots\         & & &\gate[5][50]{U} &\\
                                &             &                                 &                &                                        &               &                                 &        \ \ldots\         & & & &\\
                                &             &                                 &                &                                        &               &                                 &        \ \ldots\         & & & &\\
        \lstick[2]{$\ket{e_0}$} &             &\vqw{2}        &\setwiretype{n} & \lstick[2]{$\ket{e_0}$}                &\setwiretype{q}&\vqw{2}        &\setwiretype{n} \ \ldots\  &    &   \lstick[2]{$\ket{e_0}$}  &\setwiretype{q} &\vqw{2}        &\setwiretype{n}\\
                                &             &       &\setwiretype{n} &                                        &\setwiretype{q}&      &\setwiretype{n} \ \ldots\ &   &  &\setwiretype{q} &       &\setwiretype{n}\\
                                &             &\meter{}\arrow[from=6-3,to=7-3,arrows,  ->, shorten >=1pt, >=latex, "y_1" {anchor=south west, yshift=1pt,xshift=2pt},  at end ]           &                &                                        &               &\meter{}\arrow[from=6-7,to=7-7,arrows,  ->, shorten >=1pt, >=latex, "y_2" {anchor=south west, yshift=1pt,xshift=2pt},  at end ]           &                 &  & & &\meter{}\arrow[from=6-12,to=7-12,arrows,  ->, shorten >=1pt, >=latex, "y_t" {anchor=south west, yshift=1pt,xshift=2pt},  at end ] \\
        \lstick{}               & \qwbundle{k}         & & \cw            & \cw                                    & \cw           & &    \ \ldots\         & \cw  & \cw & \cw  & \\
    \end{quantikz} \hspace{5mm}

        \caption{Implementation of QHMM.}
        \label{fig:QHMM}
\end{figure}
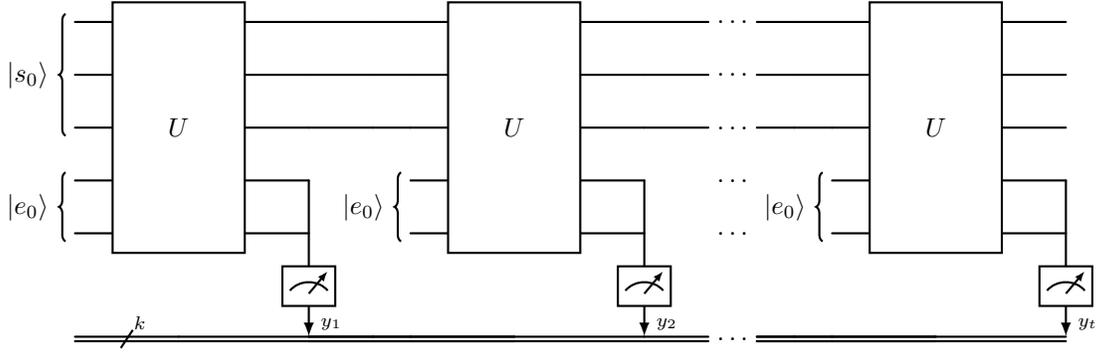

The Definition \ref{def_qhmm} of a QHMM is based on the concept of a POVM operation acting on a quantum state. The emission of the observable symbols is encoded in the operational elements (Kraus operators) of the quantum operation. 
This framework provides a convenient way to view QHMMs as channels for quantum information processing. It allows for the analysis of their informational complexity, expressive capacity, and establishes connections to stochastic process languages and the corresponding automata. 
However, this approach cannot be directly used for implementation of the QHMMs on quantum computing hardware. A critical result in quantum information theory \textemdash \ the Stinespring's representation theorem ~\cite{Stinespring1955} \textemdash \ provides approach to the physical implementation of quantum channels and correspondingly of QHMMs. According to the theorem any quantum channel can be realized as a unitary transformation on a larger system (the combined hidden state and observable systems) followed by a partial trace operation that discards the observable system. An immediate consequence of this result is the following definition of QHMMs in the unitary circuits model of computation:

\begin{definition}[Unitary quantum hidden Markov model \cite{markov2023implementation}]
    \label{def_unitary_qhmm}
    A \textit{unitary quantum} HMM $\mathbf{Q}$ over a finite alphabet of observable symbols $\Sigma$ and finite
    $N$-dimensional Hilbert space is a 6-tuple:
    \begin{equation}
    \label{QHMM_D2}
        \mathbf{Q} =  \bigl\{ \Sigma, \H_S, \H_E, U, \mathcal{M}, R_0  \bigr\}
    \end{equation}
    where
    \begin{itemize}
    \item $\Sigma$ is a finite set of $m$ observable symbols. 
    \item $\H_S$ is the Hilbert space of the hidden state system of dimension $N$. \item $\H_E$ is the Hilbert space of an auxiliary emission system with dimension $m \leq M \leq N^2$ and orthonormal basis $E=\{ \ket{e_i} \}^{M-1}_{i = 0}$.
    \item $U$ is a unitary operator defined on the bipartite Hilbert space $\H_S \otimes \H_E$.
    \item $\mathcal{M}$ is a bijective map $\mathcal{P}_m^{E} \rightarrow \Sigma$, where $\mathcal{P}_m^{E}$ is an $m$-element partition of $E$.
    \item $ R_0 = \rho_0 \otimes \ket{e_0}\bra{e_0}$ is an initial state.
\end{itemize}
\end{definition}
An implementation of QHMM where $\rho_0=\ket{s_0}\bra{s_0}$ is presented on Figure \ref{fig:QHMM}.
This unitary physical realization of a quantum channel (\textit{Definition} \ref{def_qhmm}) follows from the \textit{operator-sum representation} \cite{nielsen_chuang_2010} of the quantum operation $\mathcal{T}$ :
\begin{equation}
    \label{eqn:t_u}
    \mathcal{T}\cdot = \sum_{e \in \{ \ket{e_i} \}^{M-1}_{i = 0}} T_e \cdot T_e^{\dagger}.
\end{equation}
where the Kraus operators $\{T_e\}$ act on the hidden system states and are defined as follows:
\begin{equation}
    \label{eqn:kraus_op}
    T_e = \left( I^N \otimes \bra{e} \right) U \left( I^N \otimes \ket{e_0} \right).
\end{equation}
\noindent
The Kraus operators depend on the unitary $U$, and the arbitrarily selected orthonormal basis $\{ \ket{e} \}$ of the emission system.

    \begin{table*}[ht]
        \centering 
        \begin{tabular}{|l|l|c|c|c|c|c|c|}
            \hline
            $State$ & Direction Bias &  $0$ & 1 & 2 & 3 & $P[0|State]$ &  $P[1|State]$ \\
            \hline
            0 &Bear - Tendency Down  & 0.50 & 0.10 & 0.15  & 0.25  & 0.80 &	0.20\\
            1 &Bull - Tendency Up    & 0.10 & 0.50 & 0.25 & 0.15  & 0.20 &	0.80\\
            2 &Transition to Bear    & 0.25 & 0.15 & 0.50 & 0.10  & 0.40 &	0.60\\
            3 &Transition to Bull    & 0.15 & 0.25 & 0.10 & 0.50  & 0.60 &	0.40\\
            \hline
        \end{tabular}
            \caption{Hidden states descriptions, transition probabilities, and observation probabilities.}
        \label{tab:hmm_example}
    \end{table*}

\begin{figure*}[ht]
        \centering
        \begin{minipage}{.4\textwidth}
            \centering
            \includegraphics[width=0.7\linewidth]{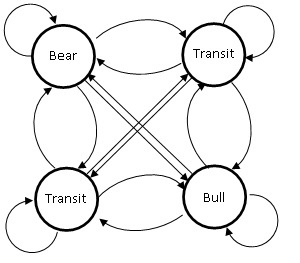}
        \end{minipage}
        \begin{minipage}{.59\textwidth}
            \centering
            \includegraphics[width=1.0\linewidth]{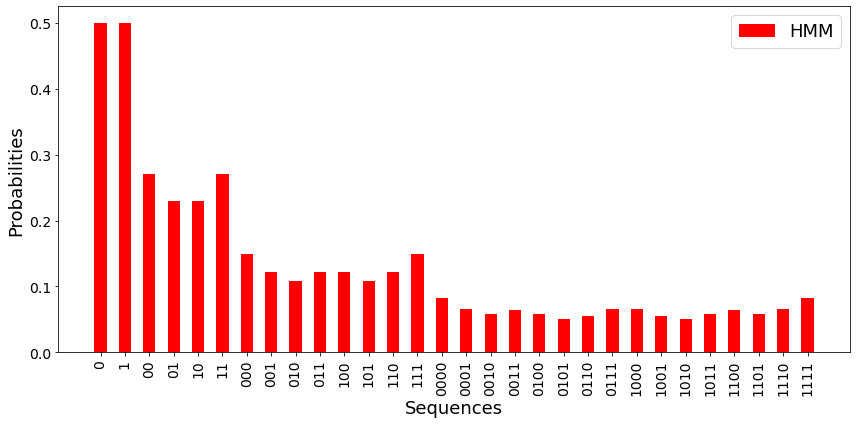}
        \end{minipage}
 
        \caption{Left: Market hidden states transition graph. Right: Distributions of observed sequences.}
        \label{fig:hmm_example}
    \end{figure*}

\begin{figure*}[ht]
    \centering
        \includegraphics[scale=0.6]{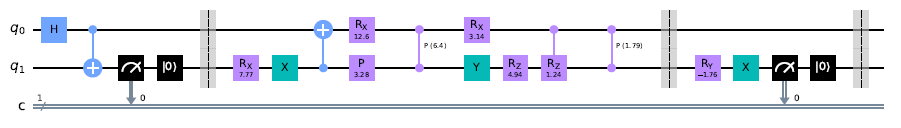}
         \caption{QHMM defining market distribution}
    \label{fig:example31fig}
\end{figure*}

\begin{figure*}[ht]
        \includegraphics[scale=1]{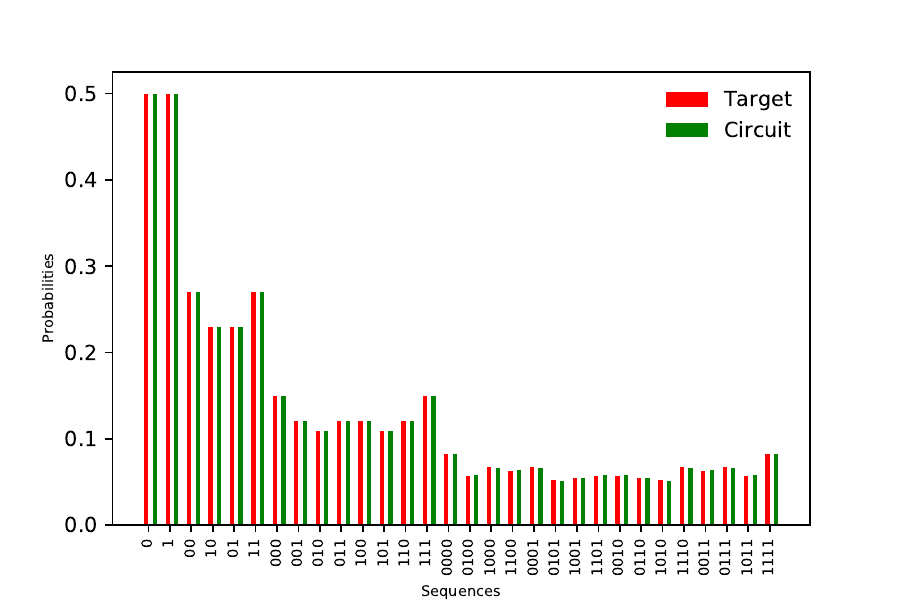}
    \caption{Targeted and Learned Market Process Language Distributions}
    \label{fig:example32fig}
\end{figure*}

\begin{example}
\label {exm: market}
 \normalfont
In \cite{markov2023implementation} we discussed a stochastic process language generated by a simple classic hidden Markov model (HMM) of directional market price movements. The market is assumed to have four hidden states and observable symbols ${0,1}$ corresponding to price move down and up. Table \ref{tab:hmm_example} provides the transition and emission probabilities of the model. The hidden states transition graph and the distributions defining the stochastic process language are presented on Figure \ref{fig:hmm_example}. In \cite{markov2023implementation} we proved that every classical HMM of $n$ states can be simulated by a QHMM (\ref{QHMM_D2}) in Hilbert space with dimension $\sqrt{n}$. Following this result we identified a generative model \textemdash \ a QHMM with one state qubit and one emission qubit (Figure \ref{fig:example31fig}) \textemdash \ which reproduces exactly the distributions of the market process language as shown on Figure \ref{fig:example32fig}. The first part of the circuit prepares a \textit{maximally mixed} initial state, the middle part performs the hidden state transition, and at the end is the measurement of the emission subsystem in a learned orthonormal basis. 
\end{example}

\section{\label{section:quantum_kernwls} Quantum Generative Kernels}
In this section, we introduce sequence similarity measures based on the quantum generative model defined by equations (\ref{qhmm}) and 
(\ref{eqn: D_M^T quantum}). We assume that a QHMM which defines the distribution set (\ref{eqn: D_M^T quantum}) of the example space is specified either as a quantum channel (\ref{qhmm}) or unitary model (\ref{QHMM_D2}).

\subsection{\label{subsubsection:predictive_kernels}Predictive Generative Kernels }
The predictive similarity measure is defined by the stochastic distance between the expected future evolutions of the sequences. Since the evolution of a Markovian process depends only on its current state, it is easy to verify that if the quantum states $\rho_1$ and $\rho_2$ defined by any two sequences $\mathbf{y^1}$ and $\mathbf{y^2}$ (\ref{seq defined state}):

$$\rho_1 = \frac{T_{\mathbf{y^1}}\rho_0}{P\bigl[ {\mathbf{y^1} \vert \rho_0 \bigr]}} $$

$$\rho_2= \frac{T_{\mathbf{y^2}}\rho_0}{P\bigl[ {\mathbf{y^2} \vert \rho_0 \bigr]}},  $$
\noindent
are the same, then the sequences $\mathbf{y^1}$ and $\mathbf{y^2}$ are considered \textit{equivalent} in the sense, that they define the same future distributions of the observables.

The following proposition generalizes this idea by proving that if two states are close in trace norm, the corresponding forward distributions of the sequences are close in total variation distance.

\begin{proposition}
\label{proposition1}
For any two states $\rho_1$ and $\rho_2$ of a QHMM $\mathbf{Q}$ the total variation distance of the observable distributions $P \bigl[ \mathbf{z}  \vert \rho_1 \bigr]$ and $P \bigl[ \mathbf{z}  \vert \rho_2 \bigr]$ for $\mathbf{z}\in\Sigma^{k}, k>0 $ is bounded by the trace distance between the states $\rho_1$ and $\rho_2$.
\end{proposition}
\begin{proof}

The similarity between the quantum states $\rho_{1}$ and $\rho_{2}$ can be estimated by their trace distance:

\begin{equation}
\label{trace_distance}
\mathcal{D}(\rho_{1},\rho_{2})= \frac{1}{2}\text{tr} |\rho_1 - \rho_2|, 
\end{equation}
where
\begin{equation*}
|\rho_i| \equiv \sqrt{\rho_i^{\dagger}\rho_i}, i=1,2
\end{equation*}

The probabilities the same sequence $\mathbf{z} \in \Sigma^{k}$ to be observed at states $\rho_{1}$ and $\rho_{2}$ respectively are (\ref{eqn:seq_probability})

\begin{equation}
P \bigl[ \mathbf{z}  \vert \rho_1 \bigr] = \operatorname{tr}(T_{\mathbf{z}} \rho_1),
\end{equation}
and
\begin{equation}
P \bigl[ \mathbf{z}  \vert \rho_2 \bigr] = \operatorname{tr}(T_{\mathbf{z}} \rho_2).
\end{equation}
\noindent
The difference between these probabilities is defined as follows:

\begin{equation}
P \bigl[ \mathbf{z}  \vert \rho_1 \bigr] - P \bigl[ \mathbf{z}  \vert \rho_2 \bigr] = \operatorname{tr}(T_{\mathbf{z}}(\rho_1 - \rho_2)).
\end{equation}
\noindent
Since the operation $T_z$ is trace non-increasing we have: 

\begin{equation}
P \bigl[ \mathbf{z}  \vert \rho_1 \bigr] - P \bigl[ \mathbf{z}  \vert \rho_2 \bigr] \leq \operatorname{tr}(\rho_1 - \rho_2).
\end{equation}
\noindent
If we assume that $P \bigl[ \mathbf{z}  \vert \rho_1 \bigr] \geq P \bigl[ \mathbf{z}  \vert \rho_2 \bigr]$ then

\begin{equation}
\label{any z }
\bigl|P \bigl[ \mathbf{z}  \vert \rho_1 \bigr] - P \bigl[ \mathbf{z}  \vert \rho_2 \bigr]\bigr| \leq 2\mathcal{D}(\rho_{1},\rho_{2}).
\end{equation}
The total variation distance between the distributions $P_1(\mathbf{z}) = P \bigl[ \mathbf{a}  \vert \rho_1 \bigr]$ and $P_2(\mathbf{z}) = P \bigl[ \mathbf{z}  \vert \rho_2 \bigr]$ is
\begin{equation}
\label{sup z}
\delta(P_1,P_2) = \sup_{\mathbf{z}\in\Sigma^{k}}\bigl|P_1(\mathbf{z}) - P _2(\mathbf{z})\bigr|
\end{equation}
Since (\ref{any z }) is valid for any $\mathbf{z}\in\Sigma^{k}$, then from (\ref{sup z}) follows:
\begin{equation}
\delta(P_1,P_2) \leq  2\mathcal{D}(\rho_{1},\rho_{2})
\end{equation}

\end{proof}

 This proposition motivates us to define the similarity between every sequence pair $\mathbf{y}^1$ and $\mathbf{y}^2$ by mapping them to the quantum states $\rho_{1}$ and $\rho_{2}$ of the model (\ref{qhmm}), and calculating the trace distance between these states. Further in this chapter we will proof that this measure of similarity pertains to the future evolution of the time series.
 
We define the predictive generative kernel as follows:
\begin{definition}[Predictive Generative Quantum Kernel]
\label{def_predictive_kernel}
Let $\mathbf{Q}$ be a Quantum Hidden Markov Model (QHMM) with domain of observable sequences $L^{\mathbf{Q}}$. For any two sequences $\mathbf{y}^1, \mathbf{y}^2 \in L^{\mathbf{Q}}$, the predictive generative quantum kernel $\kappa_p: L^{\mathbf{Q}} \times L^{\mathbf{Q}} \to \mathbb{R}$ is defined as:
\begin{equation}
\label{eq_predictive_kernel}
\kappa_p(\mathbf{y}^1, \mathbf{y}^2) = \exp\left(-\mathcal{D}\left( \phi(\mathbf{y}^1), \phi(\mathbf{y}^2) \right)\right),
\end{equation}
where $\phi: L^{\mathbf{Q}} \to \mathcal{D}(\mathcal{H})$ maps each sequence $\mathbf{y}$ to a normalized quantum state $\rho_{\mathbf{y}}$(\ref{seq defined state}):
\begin{equation}
\phi(\mathbf{y}) = \frac{T_{\mathbf{y}} \rho_0}{P[\mathbf{y} \mid \rho_0]},
\end{equation}
with $T_{\mathbf{y}}$ denoting the sequence of Kraus operators associated with $\mathbf{y}$, $\rho_0$ the initial state of the system, $P[\mathbf{y} \mid \rho_0]$ the probability of observing sequence $\mathbf{y}$ given the initial state,  and $\mathcal{D}(\rho_{\mathbf{y}}, \rho_{\mathbf{z}})$ is the trace distance between the quantum states (\ref{trace_distance}).
\end{definition}
Since the trace distance is a conditionally negative definite (CND) metric on quantum states, it follows from Schoenberg’s theorem \cite{Vapnik1998} that the function $\kappa_p$ (\ref{eq_predictive_kernel}) is a positive semi-definite (PSD) kernel.

Critical feature of the kernel is that it defines a bound of the dissimilarity of sequences with respect of their probabilistic classification maps (\ref{clsmap_p}) as it is demonstrated by the following Proposition.

\begin{proposition}[Bound on Predictive Classification Divergence]
\label{proposition:predictive_classification_kernel_bound}
Let $\kappa_p$ be the predictive generative kernel defined by (\ref{eq_predictive_kernel}). Then for any two sequences $\mathbf{y}^1, \mathbf{y}^2 \in L$ and any horizon $k > 0$, the variation distance between their predictive classification maps is bounded by the negative logarithm of the kernel:
\begin{equation}
\label{eq:kernel_bound_predictive_cls}
\sup_{c \in C} \left| p_p(\mathbf{y}^1, k, c) - p_p(\mathbf{y}^2, k, c) \right| \leq -C_k\log\left( \kappa_p(\mathbf{y}^1, \mathbf{y}^2) \right).
\end{equation}
\end{proposition}

\begin{proof}
The variation distance of the class distributions (\ref{clsmap_p}) for the sequences is defined as 

\begin{equation}
\delta(p_p(\mathbf{y}^1,k, c),p_p(\mathbf{y}^2,k, c)) = \sup_{c\in C }\bigl|p_p(\mathbf{y}^1,k, c)-p_p(\mathbf{y}^2,k, c)\bigr|
\end{equation}

\begin{equation}
\delta(p_p(\mathbf{y}^1,k, c),p_p(\mathbf{y}^2,k, c)) = \sup_{c\in C }\bigl|\sum_{\mathbf{z} \in \Sigma^k} (P[\mathbf{z} \vert \mathbf{y}^1]P[c \vert \mathbf{y^1z}]) - \sum_{\mathbf{z} \in \Sigma^k} (P[\mathbf{z} \vert \mathbf{y}^2]P[c \vert \mathbf{y^2z}]) \bigr|
\end{equation}

\begin{equation}
\delta(p_p(\mathbf{y}^1,k, c),p_p(\mathbf{y}^2,k, c)) \leq c^*\bigl|\sum_{\mathbf{z} \in \Sigma^k} (P[\mathbf{z} \vert \mathbf{y}^1]) -P[\mathbf{z} \vert \mathbf{y}^2]) \bigr|
\end{equation}

\begin{equation}
\delta(p_p(\mathbf{y}^1,k, c),p_p(\mathbf{y}^2,k, c)) \leq c^*\sum_{\mathbf{z} \in \Sigma^k} (\bigl|P[\mathbf{z} \vert \mathbf{y}^1]) -P[\mathbf{z} \vert \mathbf{y}^2]\bigr|) 
\end{equation}

\noindent
From Proposition \ref{proposition1}
\begin{equation}
\delta(p_p(\mathbf{y}^1,k, c),p_p(\mathbf{y}^2,k, c)) \leq c^* |\Sigma^k|2\mathcal{D}(\rho_{1},\rho_{2})
\end{equation}
\begin{equation}
\delta(p_p(\mathbf{y}^1,k, c),p_p(\mathbf{y}^2,k, c)) \leq -C_k\log( \kappa_p(\mathbf{y}^1,\mathbf{y}^2) ),
\end{equation}
\noindent
where $c^* \leq 1$ is a constant dependent on the particular probabilistic classification map $P[c \vert \mathbf{yz}]$ and $C_{k} =2c^* |\Sigma^k|$.

\end{proof}
Several predictive quantum kernels can be defined using alternative distance or similarity measures between quantum states, such as the Bures metric and quantum fidelity.

The Bures metric between two quantum states is defined as:
$$ \mathcal{B}(\rho_{\mathbf{y}}, \rho_{\mathbf{z}}) = 2 - 2\sqrt{F(\rho_{\mathbf{y}}, \rho_{\mathbf{z}})}, $$
\noindent
where the quantum fidelity $F(\rho_{\mathbf{y}}, \rho_{\mathbf{z}})$ is defined as:
\[ F(\rho_{\mathbf{y}}, \rho_{\mathbf{z}}) = \left(\text{Tr}\left(\sqrt{\sqrt{\rho_{\mathbf{y}}} \rho_{\mathbf{z}} \sqrt{\rho_{\mathbf{y}}}}\right)\right)^2. \]
Using the Bures metric, we can define a valid predictive kernel via negative exponentiation:
\begin{equation}
\label{bures_predictive_kernel}
\kappa_p(\mathbf{y},\mathbf{z}) = \exp\left(-\mathcal{B}\bigl( \phi(\mathbf{y}), \phi(\mathbf{z}) \bigr)\right), \quad \mathbf{y},\mathbf{z} \in L^{\mathbf{Q}}.
\end{equation}
Since the Bures metric is conditionally negative definite (CND), the kernel defined in \eqref{bures_predictive_kernel} is positive semi-definite (PSD) by Schoenberg’s theorem.

\noindent
We can also define a predictive kernel directly from fidelity as follows:
\begin{equation}
\label{fidelity_predictive_kernel}
\kappa_p(\mathbf{y},\mathbf{z}) = F(\rho_{\mathbf{y}}, \rho_{\mathbf{z}}).
\end{equation}
Fidelity is a symmetric, bounded, and positive-definite similarity measure, and thus the function \eqref{fidelity_predictive_kernel} defines a valid PSD kernel.

\subsection{\label{subsubsection:structural_kernels}Structural Generative Kernels }
The classification maps of the structural classification tasks (\ref{clsmap_s}) depend on the probabilistic process that generates the sequences. 
The generative process is defined by the evolution of model's quantum state.
For any example $\mathbf{y} \in L^{Q}, \mathbf{y} = y_1y_2\cdots y_n, y_i \in \Sigma, i \in [1,n]$ a \textit{generating state sequence} of $\mathbf{y}$ is defined as (\ref{seq defined state}):
\begin{equation}
\label{generating state sequence }
 \left\{ \rho_i = \frac{T_{y_{i}}\rho_{i-1}}{tr(T_{y_{i}}\rho_{i-1})}:i \in [1,n]  \right\}. 
\end{equation}
\noindent
We introduce a kernel that estimates the structural similarity of examples based on the divergence of appropriate statistics derived from the generating state sequences. The structure of each sequence depends on the level of involvement of model's state in the generative process: if a state participates more often, its impact is stronger. To account for this dependency, the structural kernel maps each example to the expectation of its generating states. Then, the structural similarity of sequences is measured by the divergence of the expectations of their generating states.
The map of examples to the expectations of quantum states is defined using (\ref{generating state sequence }) as follows: 
$$\phi_s(\mathbf{y})=  \hat{\rho}_{\mathbf{y}}=\frac{1}{n}\sum_{i\in[1,n]}\rho_i, $$
\noindent
where $\mathbf{y} = y_1y_2\cdots y_n$. 
\noindent
It is easy to demonstrate, that the average of a set of density matrices is a density matrix. Therefore, the expectation $\hat{\rho}$ is a quantum state and we can define a valid structural quantum kernel using the trace distance as follows:

\begin{equation}
\label{structural_kernel}
\kappa_s(\mathbf{y},\mathbf{z}) = \exp\left(-\mathcal{D}\bigl (  \phi_s(\mathbf{y}), \phi_s(\mathbf{z}) \bigr )\right), \mathbf{y},\mathbf{z} \in L^Q 
\end{equation}
\noindent
A valid structural kernel can also be defined using the Bures metric:

\begin{equation}
\label{structural_kernel_b}
\kappa_s(\mathbf{y},\mathbf{z}) = \exp\left(-\mathcal{B}\bigl (  \phi_s(\mathbf{y}), \phi_s(\mathbf{z}) \bigr )\right), \mathbf{y},\mathbf{z} \in L^Q 
\end{equation}

\section{\label{section:empirical_evaluation} Empirical Evaluation of Quantum Generative Kernels} 
In this section, we use the stochastic process language discussed in Example \ref{exm: market} and the corresponding generative QHMM (Figure \ref{fig:example31fig}) to empirically investigate the behaviour of the introduced kernels in several contexts: dimension of the quantum Hilbert space, type of the classification task, and type of the stochastic divergence measures of the quantum states. The QHMM discussed in Example \ref{exm: market} is minimal, with dimension of the state Hilbert space $N=2$ (\ref{def_unitary_qhmm}). To study the impact of the size of the quantum Hilbert space on kernels behaviours we learned and investigated two larger models of the same stochastic process language with dimensions $N=4$ and $N=16$ correspondingly. The components of these models: initial state preparation, transition algorithm, and observable emission in circuit quantum computing model are presented in appendix \ref{Appendix Higher Dim}.

\subsection{Quantum Feature Space Dimension Impact}
\begin{figure*}[ht]

    \begin{minipage}{.5\textwidth}
        \centering
        \includegraphics[trim = 0.0mm 0.0mm 0.0mm 0.0mm, clip,width=1.\linewidth]{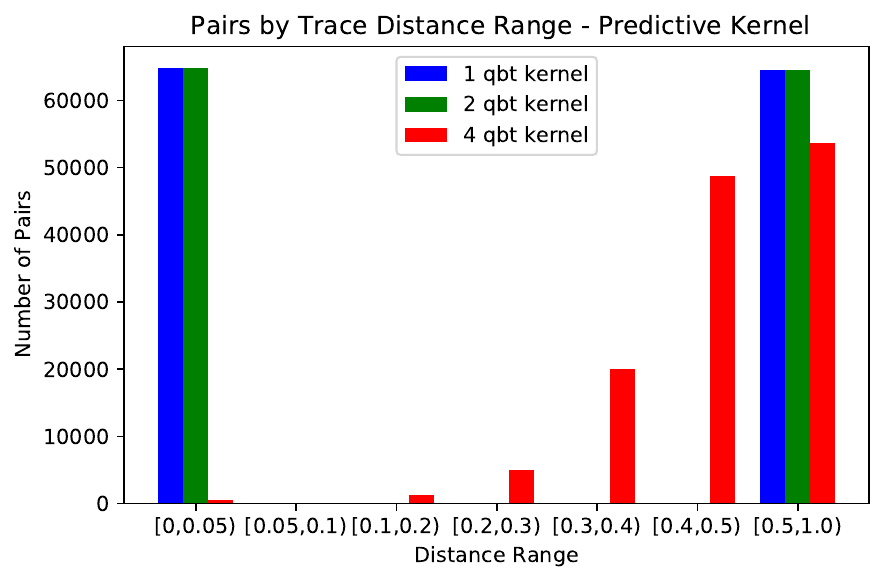}
        
        
    \end{minipage}
    \begin{minipage}{.5\textwidth}
        \centering
        
        \includegraphics[trim = 0.0mm 0.0mm 0.0mm 0.0mm, clip,width=1.\linewidth]{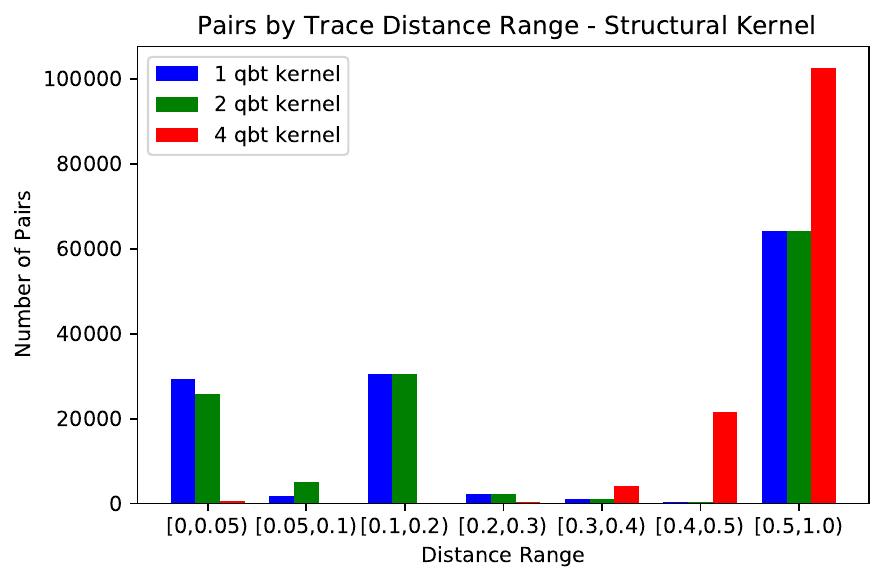}
        
    \end{minipage}
    \caption{Number of Pairs by 
    Distance  - Trace Metric }
    \label{fig:Trace_Pred_Struct}
\end{figure*}

\begin{figure}[ht]
    \begin{minipage}{.5\textwidth}
        \centering
        \includegraphics[trim = 0.0mm 0.0mm 0.0mm 0.0mm, clip,width=1.\linewidth]{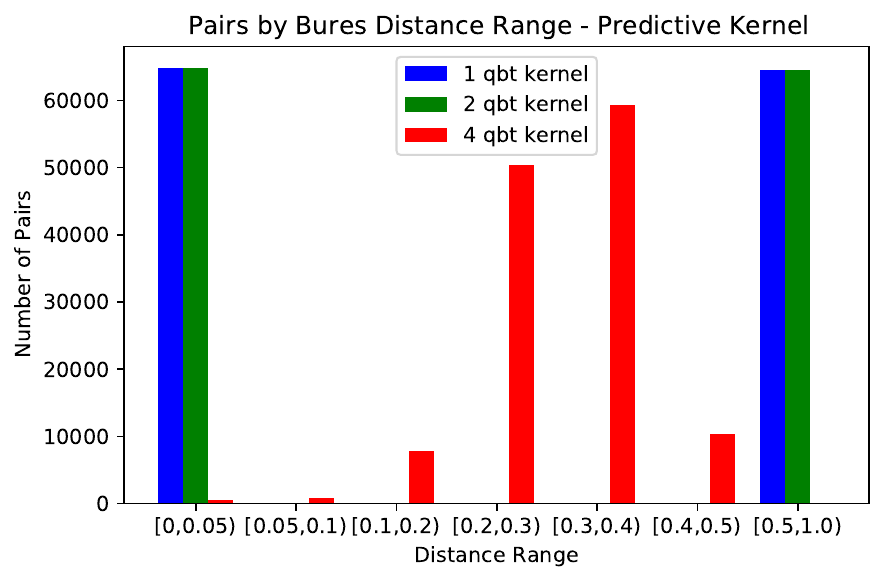}
        
        
    \end{minipage}
    \begin{minipage}{.5\textwidth}
        \label{Trace_Pred_Struct}
        \centering
        
        \includegraphics[trim = 0.0mm 0.0mm 0.0mm 0.0mm, clip,width=1.\linewidth]{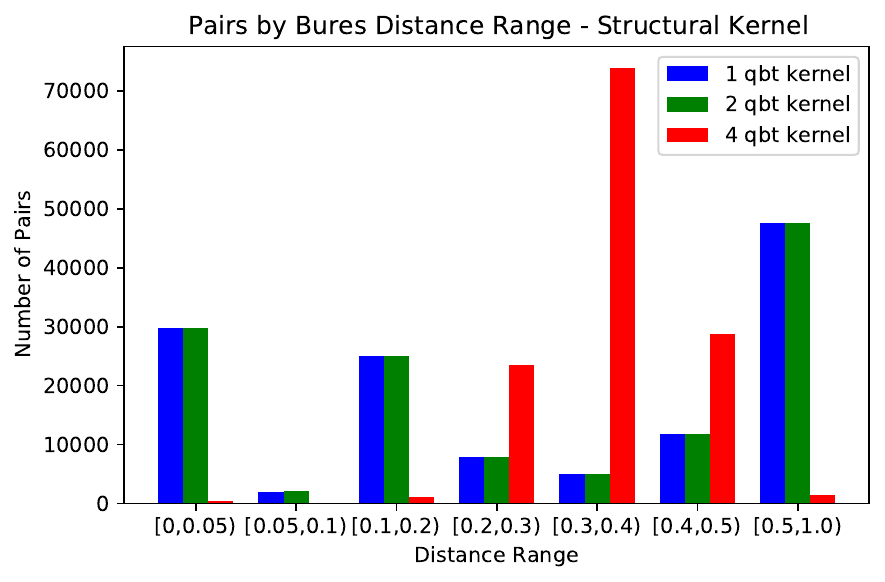}
        
        
    \end{minipage}
    \caption{Number of Pairs by Distance - Bures Metric}
    \label{fig:Bures_Pred_Struct}
\end{figure}

It is important to demonstrate how the introduced kernels support the expectation that mapping the original domain into a high-dimensional Hilbert space enhances the separability of examples. For QHMMs of the Market example with Hilbert space dimensions 2, 4, and 16, we measured the distributions of the distances between examples induced by the kernels. The results are presented in Figures \ref{fig:Trace_Pred_Struct} and \ref{fig:Bures_Pred_Struct}. 
If a kernel maps a significant number of pairs to short distance ranges, indicating they are not well distinguishable, this behavior could pose a potential problem for kernel learning methods to perform effectively in nonlinear environments. On the other hand, if a kernel assigns significant distances to most pairs of examples, it might be more successful in challenging learning tasks.

The initial observation reveals that the Predictive kernels in low-dimensional Hilbert spaces struggle to distinguish half of the examples while effectively separating the other half. On the other hand, the Structural kernels on 1 and 2 Qubits display better performance by effectively separating a significant portion of the examples. Furthermore, it is evident that the Trace metric yields better separation than the Bures metric. Notably, the 4 Qubit kernel consistently performs well across all scenarios, as the Structural kernel with the Trace metric achieves perfect separation of all pairs of examples. This behavior underscores the direct impact of increasing the dimension of the Hilbert space of the QHMM on the kernel's performance.

\subsection{Kernel-Induced Distances with Respect to Classification Tasks}
\begin{figure}[ht]
    \begin{minipage}{.5\textwidth}
        \centering
        \includegraphics[trim = 0.0mm 0.0mm 0.0mm 0.0mm, clip,width=1.\linewidth]{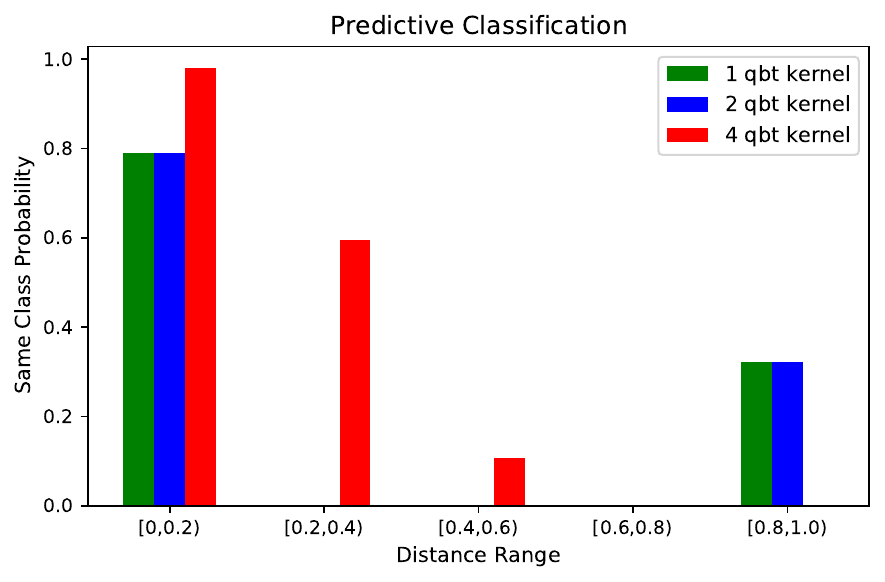}
        
        
    \end{minipage}
    \begin{minipage}{.5\textwidth}
        \centering
        
        \includegraphics[trim = 0.0mm 0.0mm 0.0mm 0.0mm, clip,width=1.\linewidth]{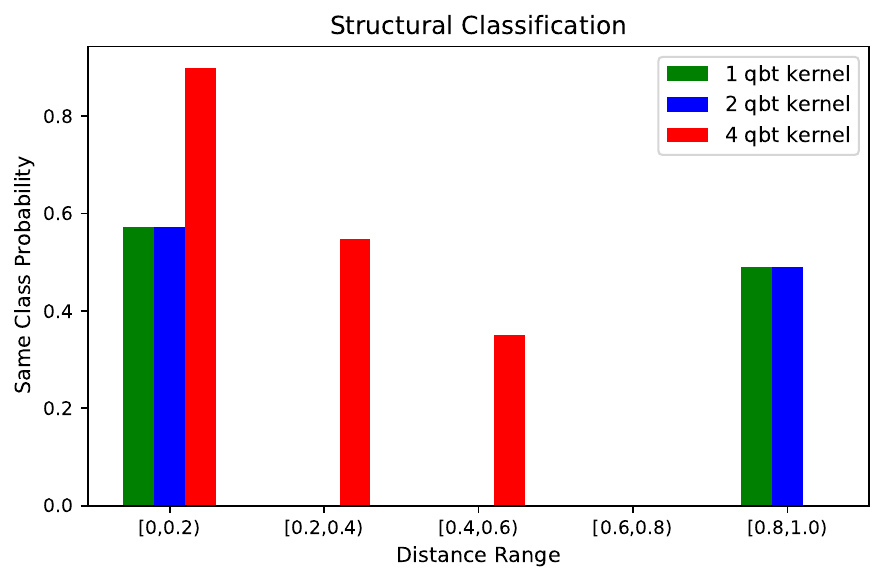}
        
    \end{minipage}
    \caption{Probability of Examples Belonging to the Same Class - Predictive Kernel, Bures Metric}
\label{fig:Bures_Predictive_Kernel_Class}
\end{figure}

\begin{figure}[ht]
    \begin{minipage}{.5\textwidth}
        \centering
        \includegraphics[trim = 0.0mm 0.0mm 0.0mm 0.0mm, clip,width=1.\linewidth]{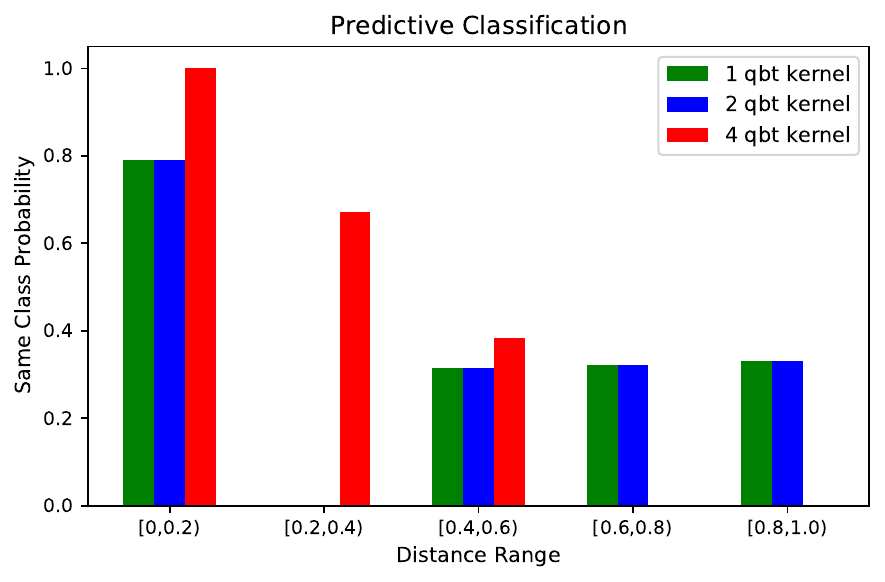}
        
        
    \end{minipage}
    \begin{minipage}{.5\textwidth}
        \centering
        
        \includegraphics[trim = 0.0mm 0.0mm 0.0mm 0.0mm, clip,width=1.\linewidth]{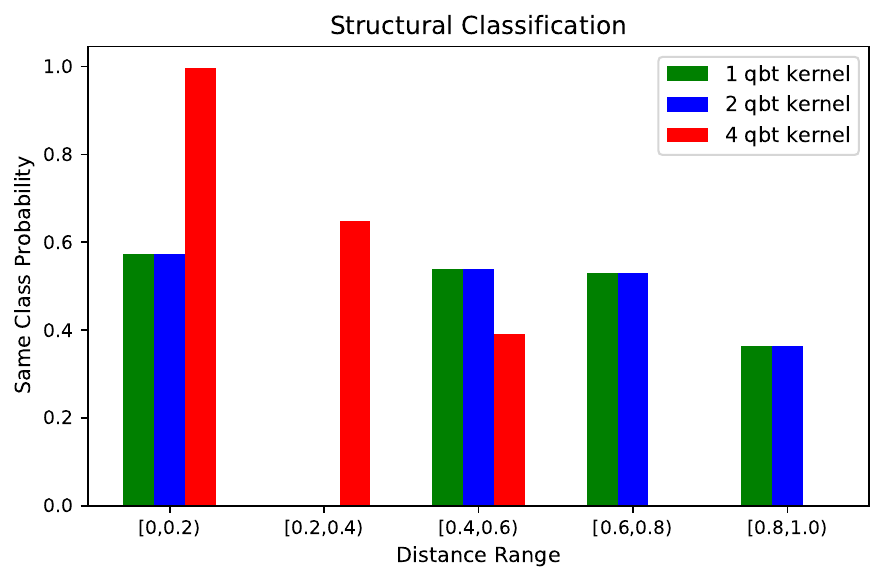}
        
    \end{minipage}
    \caption{Probability of Examples Belonging to the Same Class - Structural Kernel, Bures Metric}
\label{fig:Bures_Struct_Kernel_Class}
\end{figure}

We demonstrate that distances calculated by the structural and predictive kernels are correlated with the classes of corresponding examples. Specifically, examples at shorter distances tend to belong to the same class. We will consider simple structural and predictive classification tasks as defined in appendix \ref{Appendix Classifiaction tasks}. 
On figures \ref{fig:Bures_Predictive_Kernel_Class} and \ref{fig:Bures_Struct_Kernel_Class} it is evident that the distances defined by the 4-qubit kernels exhibit an inverse linear relationship with the probability of examples belonging to the same class. The stronger correlation is demonstrated by the 4-qubit predictive kernel in the predictive classification task. Conversely, the performance of the 1-qubit and 2-qubit structural and predictive kernels is the poorest in the structural classification task. Similar results, presented in appendix \ref{Appendix Kernel Induced Distances}, were obtained for the trace metric. 
Ultimately, the dimensionality of the quantum feature space significantly impacts kernels' performance.

\subsection{Quantum Kernels vs Classical Kernels}
To compare the performance of the proposed kernels against classical ones, we use the classification tasks described in appendix \ref{Appendix Classifiaction tasks}. Three common algorithms \cite{scikit-learn} - random forest classifier, support vector classifier, and kernelized k-nearest neighbours classifier were applied to the classification tasks. The random forest classifier was used as a bench mark and the other were implemented with classical and quantum kernels. As classical kernel was the radial basis function (RBF) kernel defined as follows \cite{Vapnik1998}:
$$K(\mathbf{y}^1, \mathbf{y}^2) = \operatorname{exp}\left(-\frac{ \left\lVert (\mathbf{y}^1 - \mathbf{y}^2) \right\rVert^2 }{2\sigma^2}\right), $$
where $\left\lVert (\mathbf{y}^1 - \mathbf{y}^2) \right\rVert$ is the Euclidean distance between the sequences, and $\sigma$ is a parameter.

In all structural and predictive task scenarios, the quantum kernels exhibited superior performance compared to their classical counterparts and the RFS benchmark. The algorithms were tested on 500 samples split to training and testing at $50\%$ using Trace metric (\ref{eq_predictive_kernel}) for the kernels. As it can be seen in Table \ref{tab:predictive_comparison} and Table \ref{tab:structural_comparison} in both classification tasks the quantum kernels outperformed their classical counterparts as the SVC with quantum kernel outperformed the RFS benchmark algorithm as well. 

Similar performance were observed with kernels using Bures metric (\ref{bures_predictive_kernel}), (\ref{structural_kernel_b}) as the results are presented in appendix \ref{Appendix Kernels Performance - Distance Metric Bures}.

    \begin{table*}[ht]
        \centering 
        \begin{tabular}{|c|c|c|c|c|c|}
            \hline
            Classifier & Kernel &  In Sample  & CI & Out Sample  & CI \\
            \hline
            RFS & N/A       & 0.998 & 0.996 - 0.999 & 0.968  &  0.966 - 0.971\\
            SVC & Classical & 0.951 & 0.947 - 0.955 & 0.916  &  0.910 - 0.922\\
            SVC & Quantum   & 0.999 & 0.999 - 1.000 & 0.980  &  0.977 - 0.981\\
            k-NN& Classical & 0.971 & 0.969 - 0.973 & 0.916  &  0.913 - 0.919 \\
            k-NN& Quantum   & 0.993 & 0.992 - 0.994 & 0.980  &  0.978 - 0.982 \\
            \hline
        \end{tabular}
            \caption{Kernels Performance by Recognizer Accuracy: Predictive Task, Predictive Kernel, Distance Metric- Trace, CI=95\%, Kernel Type RBF}
        \label{tab:predictive_comparison}
    \end{table*}

    \begin{table*}[ht]
        \centering 
        \begin{tabular}{|c|c|c|c|c|c|}
            \hline
            Classifier & Kernel &  In Sample  & CI & Out Sample  & CI \\
            \hline
            RFS & N/A       & 0.985 & 0.984 - 0.987 & 0.824  &  0.819 - 0.829\\
            SVC & Classical & 0.962 & 0.959 - 0.965 & 0.903  &  0.896 - 0.909\\
            SVC & Quantum   & 1.000 &  -  & 0.949  &  0.946 - 0.952\\
            k-NN& Classical & 0.863 & 0.857 - 0.869 & 0.720  &  0.715 - 0.723 \\
            k-NN& Quantum   & 0.950 & 0.948 - 0.953 & 0.893  &  0.889 - 0.897 \\            \hline
        \end{tabular}
            \caption{Kernels Performance by Recognizer Accuracy: Structural Task, Structural Kernel, Distance Metric- Trace, CI=95\%, Kernel Type RBF}
        \label{tab:structural_comparison}
    \end{table*}
    
\section{\label{section:quantum_kernels_circuits} Quantum Kernels in Quantum Circuits Computing Model}

For each sequence $\textbf{a} \in {\Sigma}^{*}$ a quantum generative model $Q$ (\ref{qhmm}) defines a probability 
 $P \bigl[ \textbf{a}  \vert G \bigr] $, and hence $G$ defines distribution on all continuations $\textbf{b}, \vert \textbf{b} \vert = t $ of the sequence: 

$$ D_{t}^{\textbf{a}} = \bigl\{ P\bigl[ \textbf{b} \vert \textbf{a}, G  \bigr] = P\bigl[ \textbf{a}\textbf{b} \vert G \bigr]  : \textbf{b} \in \Sigma^{t} \bigr\} $$
In many cases the class of a sequence depends on the current state and the probability of its continuations. For example, in a financial time series of price movements the class of a sequence can be 1 if the next movement is expected to be ``UP". In an English sentence the class of a phrase is ``subject" if the expected next word is a verb. In these cases we classify the sequence by the distributions of their continuations. Therefore it is natural to assume that if the induced future distributions of two sequences are close, the similarity of these sequences is high.
Let's formalize this intuitive explanation. Let $\mathbf{a} = a_1 \ldots a_k $ and $\mathbf{b} = b_1 \ldots b_l $ are the sequences we want to compare. The model $Q$ defines the following end states for each of them:

$$\rho_{\mathbf{a}} = T_{a_k} \ldots T_{a_1} \rho_0$$
$$\rho_{\mathbf{b}} = T_{b_l} \ldots T_{b_1} \rho_0$$.

The use of quantum computing for non-linear mapping into a higher-dimensional vector space for non-linear classification was discussed for almost a decade in different sources, e.g. \cite{Rebentrost2014Quantum,Chatterjee2017Generalized,schuld2019quantum,park2020practical,rastunkov2022boosting}.

In the field of quantum machine learning a quantum kernel is often defined as the inner product between two data-encoding feature vectors $\rho_{\mathbf{a}}$, $\rho_{\mathbf{b}}$ \cite{schuld2021machine}:

$$k_{1,2}=Tr\{\rho_{\mathbf{a}}\rho_{\mathbf{b}}\}$$

This definition works well, when feature maps are implemented by unitary operations on quantum circuits. In quantum computing there is a vital need to verify and characterize quantum states, where the fidelity is an important and useful similarity measure. There are a number of proposals for estimating the mixed state fidelity on a quantum computer, which overcome hardness limitations of classical algorithms. One proposal is a variational quantum algorithm for low-rank fidelity estimation \cite{cerezo2020variational}. Another proposal with exponential speedup \cite{wang2022quantum} relies on state purifications being supplied by an oracle. Other variational algorithms have been proposed for the fidelity and trace distance \cite{chen2021variational}. A number of quantum measures of distinguishability are discussed in \cite{fuchs1996distinguishability}. Special consideration should be given to working with mixed states. Classical fidelity measure was introduced by Richard Jozsa \cite{jozsa1994fidelity}. \cite{liang2019quantum} gives most recent review of mixed states fidelity measures.

In the case of more general feature maps producing mixed states we will use definition \ref{def_predictive_kernel} and equation (\ref{eq_predictive_kernel}).

\begin{figure}[ht]
        \centering
        \includegraphics[trim = 85.0mm 20.0mm 65.0mm 20.0mm, clip,width=1.\linewidth]{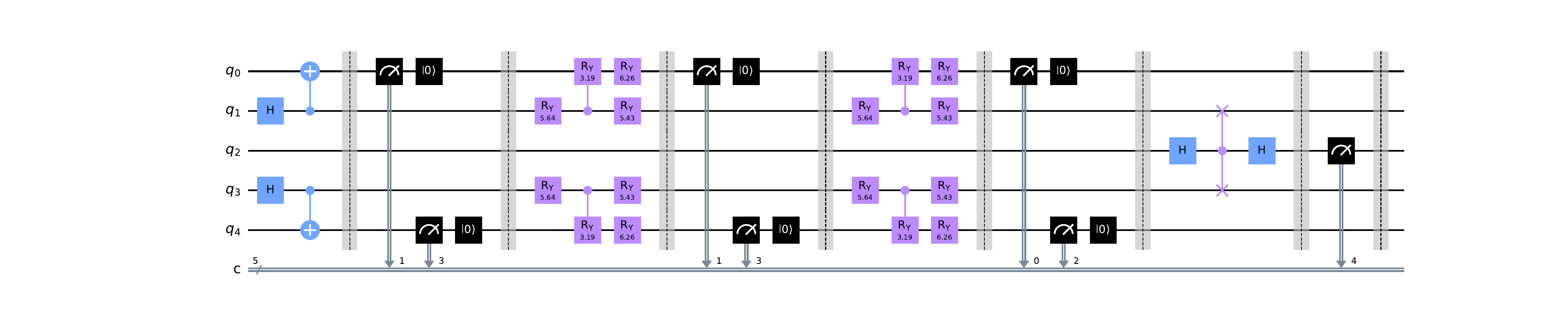}
        \caption{Implementation of SWAP test for market QHMM}
        \label{fig:marketSWAPcircuit}
\end{figure}

\begin{figure}[htbp]
    \centering
    \begin{minipage}{.5\textwidth}
        \begin{minipage}{1.\textwidth}
            \raggedright{X basis }
            \begin{quantikz}
                \lstick{$\ket{\psi}$} & \gate[wires=1]{H} & \meter{} & \qw\\
            \end{quantikz}
        \end{minipage}
        \begin{minipage}{1.\textwidth}
            \raggedright{Y basis }
            \begin{quantikz}
                \lstick{$\ket{\psi}$} & \gate[wires=1]{S^\dagger} & \gate[wires=1]{H} & \qw & \meter{} & \qw\\
            \end{quantikz}
        \end{minipage}
        \begin{minipage}{1.\textwidth}
            \raggedright{Z basis }
            \begin{quantikz}
                \lstick{$\ket{\psi}$} & \meter{} & \qw\\
            \end{quantikz}
        \end{minipage}

    \end{minipage}
    \caption{Measurements of the single qubit state in X, Y and Z bases}
    \label{fig:meas_xyz_basis}
\end{figure}
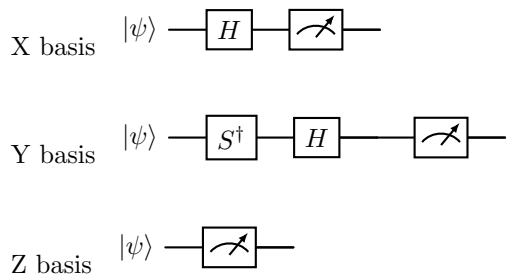

\begin{figure}[ht]
        \centering
        \includegraphics[trim = 85.0mm 20.0mm 65.0mm 20.0mm, clip,width=1.\linewidth]{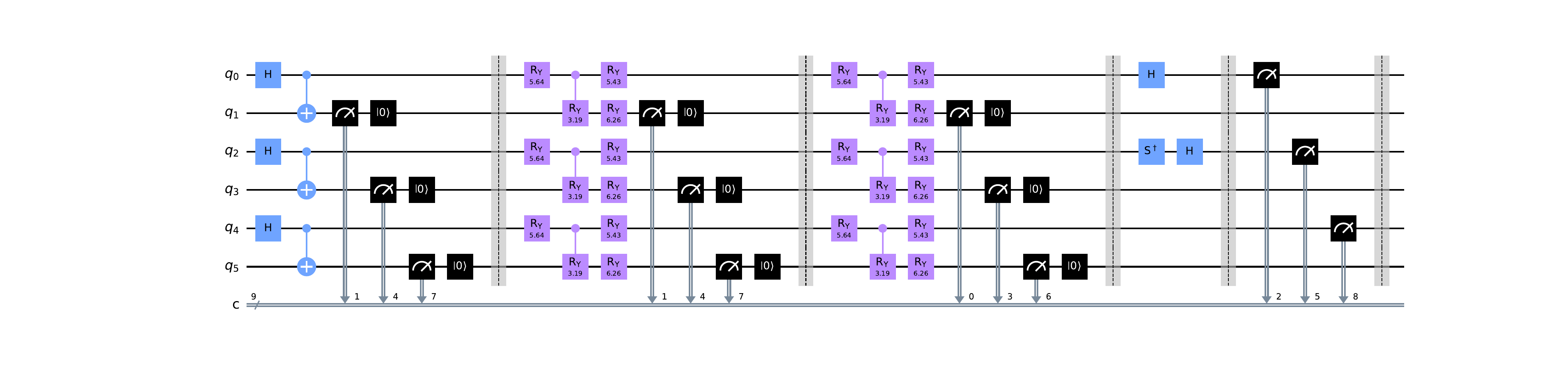}
        \caption{Implementation of projected kernel for market QHMM circuit}
        \label{fig:marketProjCircuit2x1}
\end{figure}

\begin{figure}[!ht]
    \centering
    \begin{minipage}[t]{.49\textwidth}
        \centering
        \includegraphics[trim = 15.0mm 20.0mm 15.0mm 20.0mm, clip,width=1.\linewidth]{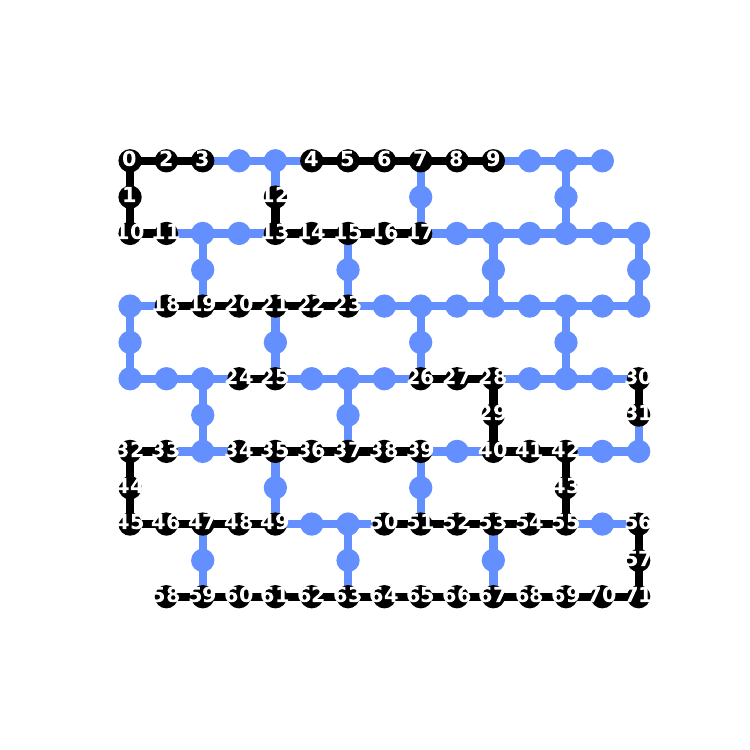}
        \caption{Initial qubit layout on \textit{ibm\_nazca} device for projected kernel approach}
        \label{fig:ibm_nazca_layout}
    \end{minipage}%
    \hfill
    \begin{minipage}[t]{.49\textwidth}
        \centering
        \includegraphics[trim = 15.0mm 10.0mm 20.0mm 1.0mm, clip,width=1.\linewidth]{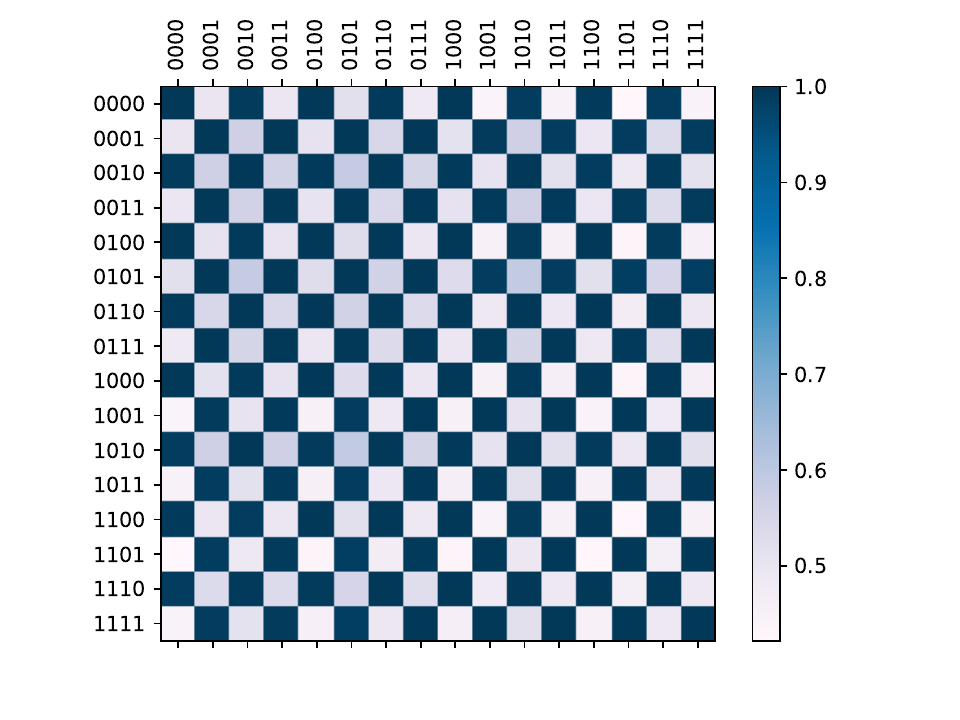}
        \caption{Color bar plot of the kernel matrix estimated on (\textit{ibm\_nazca}) device using projected kernels approach}
        \label{fig:ibm_nazcaKernelMatrixLength4}
    \end{minipage}
\end{figure}

\begin{figure}[ht]
        \centering
        \includegraphics[width=1.\linewidth]{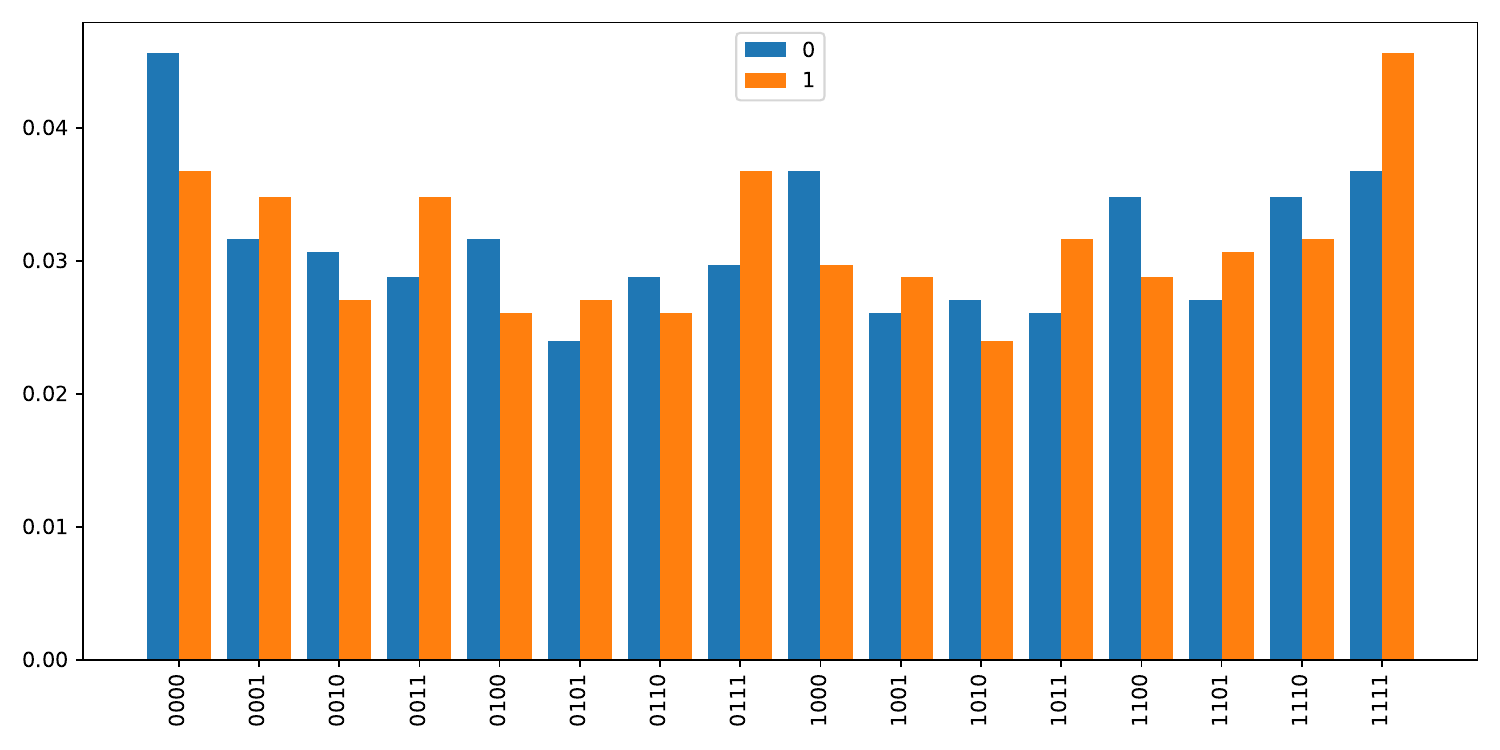}
        \caption{Next Symbol Distributions for $4$-symbol Sequences}
        \label{fig:market_hist_target_split}
\end{figure}

In this section we will construct a kernel matrix for all pairs of sequences. One possible approach to implement this on the hardware is to use SWAP test \cite{Buhrman2001Quantum} as shown on \fig{marketSWAPcircuit}. Qubits $q_0$, $q_1$ and $q_3$, $q_4$ on \fig{marketSWAPcircuit} implement two QHMMs, $q_2$ is used for fidelity calculation. Maximally mixed states are prepared for both QHMMs leveraging classical registers $1$ and $3$. Then the circuit implements two steps of QHMM generator. Classical registers $1$ and $0$ capture the symbols of the first sequence, registers $3$ and $2$ of the second. Finally, the probability of measuring $0$ on qubit $q_2$ captured on classical register $4$ can be converted to the squared fidelity \cite{Buhrman2001Quantum,havlivcek2019supervised} as 

$$\vert \langle{\phi} \vert \psi \rangle \vert^{2} = 2 \times (-0.5 + r_{0}\times R^{-1})$$

\noindent
where $\phi$ and $\psi$ are state vectors of respective state systems of both QHMMs, $r_{0}$ is the number of observed zero bit-strings on qubit $q_2$ and $R$ is the number of shots. The number of shots required is usually estimated as $O(n)$, where $n$ is the number of states. In our case, $n=2^{2t + 1}$, where $t$ is the length of the sequence. Unfortunately, for sequences of length 7-8 this requires about 1 million shots or more. Another downside is that this approach assumes calculating kernel element for pure states. In general, we can be working with mixed states. A better approach is to use projected kernels \cite{Huang2021Power}. It is well-known \cite{nielsen_chuang_2010,J__2022Quantum} that a density matrix of a single qubit can be written as 
$$\rho=\frac{1}{2} \left[1+\sum_{k}{r_k\sigma_k}\right]$$
where $\sigma_k$ are the Pauli matrices and $r_k$ is a real, three component vector. $\rho$ can be reconstructed from the three measurements results 
$$r_k=\left\langle\sigma_k\right\rangle=tr(\rho\sigma_k)$$.

The circuit implementation \cite{J__2022Quantum} is shown on \fig{meas_xyz_basis}. We can implement projected kernel for the market QHMM circuit as shown on \fig{marketProjCircuit2x1}. Finally, the kernel element is constructed from two density matrices as $k_{1,2}=e^{-\gamma \lVert \rho_1 - \rho_2 \rVert^2}$, where $\lVert . \rVert$ is a Frobenius norm and $\gamma$ is a hyperparameter (for now, we set $\gamma=1$). In order to boost the effective number of shots we propose to run multiple circuits simultaneously in parallel on a single chip by combining them on a single circuit (multi-programming \cite{das2019case,liu2021qucloud,niu2023multi}). With this in mind, the initial qubit layout on \textit{IBM Nazca} device is shown on \fig{ibm_nazca_layout}, where we effectively use $72$ qubits. Color bar plot of the kernel matrix is shown on \fig{ibm_nazcaKernelMatrixLength4}. We see a perfect separation between classes. In this case the main system required only $1$ qubit. In case, we have systems of many qubits we can use 1-reduced density matrix (1-RDM) or N-RDM approximations.

The checkerboard pattern is not surprising. Let's plot $5$-symbol sequences on a histogram and split counts for the fifth symbol between two bars, one for $0$ and one for $1$ (see \fig{market_hist_target_split}). This way we can see, which is the most likely suffix for $4$-symbol prefix. $0$ and $1$ alternate as we go from $0000$ to $1111$.

\section{\label{section:conclusion} Conclusions and Outlook}
In this article, we introduced a generative quantum computing approach to designing similarity measures and associated kernels for the classification of stochastic symbolic time series. The proposed kernels are built upon a novel class of quantum generative models known as quantum hidden Markov models. We provided theoretical justifications of the efficacy of the kernels in predictive classification tasks. Extensive simulation experiments have demonstrated the scalable and discriminative performance of the kernels in high-dimensional Hilbert spaces. Furthermore, we have devised and implemented the kernels in the quantum circuits computing model and successfully conducted experiments on quantum hardware. Finally, we compared the performance of the introduced quantum kernels to their classical counterparts on simple predictive and structural classification tasks, where the quantum kernels showed clear superiority.
The proposed approach introduces novel quantum generative modeling techniques for designing hybrid classical-quantum algorithms for clustering, classification, and regression tasks in the domain of symbol sequences and time series.
An active research direction involves applying the proposed kernels to the classification of high-frequency time series in the field of quantitative finance.
Success in this challenging domain will require extending the underlying quantum stochastic model to encompass the generation of fractional and other non-Markovian stochastic processes.

\section*{Acknowledgements}

The authors thank Amol Deshmukh for fruitful discussions about this work.

The views expressed in this article are those of the authors and do not represent the views of Wells Fargo. This article is for informational purposes only. Nothing contained in this article should be construed as investment advice. Wells Fargo makes no express or implied warranties and expressly disclaims all legal, tax, and accounting implications related to this article.

\typeout{}
\bibliographystyle{ieeetr}
\bibliography{main}

\begin{thebibliography}{10}

\bibitem{LTW01}
H.~Shimodaira, K.-i. Noma, M.~Nakai, and S.~Sagayama, ``Dynamic time-alignment kernel in support vector machine,'' in {\em Advances in Neural Information Processing Systems} (T.~Dietterich, S.~Becker, and Z.~Ghahramani, eds.), vol.~14, MIT Press, 2001.

\bibitem{Pearson}
Y.~Zheng, Q.~Liu, E.~Chen, Y.~Ge, and J.~L. Zhao, ``Time series classification using multi-channels deep convolutional neural networks,'' in {\em Web-Age Information Management} (F.~Li, G.~Li, S.-w. Hwang, B.~Yao, and Z.~Zhang, eds.), (Cham), pp.~298--310, Springer International Publishing, 2014.

\bibitem{Spearman}
J.~Ye, C.~Xiao, R.~M. Esteves, and C.~Rong, ``Time series similarity evaluation based on spearman's correlation coefficients and distance measures,'' in {\em Cloud Computing and Big Data} (W.~Qiang, X.~Zheng, and C.-H. Hsu, eds.), (Cham), pp.~319--331, Springer International Publishing, 2015.

\bibitem{TSClustering}
S.~Aghabozorgi, A.~{Seyed Shirkhorshidi}, and T.~{Ying Wah}, ``Time-series clustering – a decade review,'' {\em Information Systems}, vol.~53, pp.~16--38, 2015.

\bibitem{NIPS2001_a869ccbc}
H.~Shimodaira, K.-i. Noma, M.~Nakai, and S.~Sagayama, ``Dynamic time-alignment kernel in support vector machine,'' in {\em Advances in Neural Information Processing Systems} (T.~Dietterich, S.~Becker, and Z.~Ghahramani, eds.), vol.~14, MIT Press, 2001.

\bibitem{Edit}
L.~Chen and R.~Ng, ``- on the marriage of lp-norms and edit distance,'' in {\em Proceedings 2004 VLDB Conference} (M.~A. Nascimento, M.~T. Özsu, D.~Kossmann, R.~J. Miller, J.~A. Blakeley, and B.~Schiefer, eds.), pp.~792--803, St Louis: Morgan Kaufmann, 2004.

\bibitem{MSM}
A.~Stefan, V.~Athitsos, and G.~Das, ``The move-split-merge metric for time series,'' {\em IEEE Transactions on Knowledge and Data Engineering}, vol.~25, no.~6, pp.~1425--1438, 2013.

\bibitem{PLA}
H.~Li, C.~Guo, and W.~Qiu, ``Similarity measure based on piecewise linear approximation and derivative dynamic time warping for time series mining,'' {\em Expert Systems with Applications}, vol.~38, no.~12, pp.~14732--14743, 2011.

\bibitem{Lin03}
J.~Lin, E.~Keogh, S.~Lonardi, and B.~Chiu, ``A symbolic representation of time series, with implications for streaming algorithms,'' in {\em Proceedings of the 8th ACM SIGMOD Workshop on Research Issues in Data Mining and Knowledge Discovery, DMKD '03}, pp.~2--11, 06 2003.

\bibitem{lines2015time}
J.~Lines and A.~Bagnall, ``Time series classification with ensembles of elastic distance measures,'' {\em Data Mining and Knowledge Discovery}, vol.~29, pp.~565--592, 2015.

\bibitem{ye2011time}
L.~Ye and E.~Keogh, ``Time series shapelets: a novel technique that allows accurate, interpretable and fast classification,'' {\em Data Mining and Knowledge Discovery}, vol.~22, pp.~149--182, 2011.

\bibitem{scholkopf2001learning}
B.~Sch{\"o}lkopf and A.~J. Smola, {\em Learning with Kernels: Support Vector Machines, Regularization, Optimization, and Beyond}.
\newblock MIT press Cambridge, 2001.

\bibitem{mercer1909functions}
J.~Mercer, ``Functions of positive and negativetypeand their connection with theory ofintegral equations,'' {\em Philosophical Trinsoctions of Royal Society}, pp.~4--415, 1909.

\bibitem{Vapnik1998}
V.~N. Vapnik, {\em Statistical Learning Theory}.
\newblock John Wiley \& Sons, Inc, 1998.

\bibitem{boser1992training}
B.~Boser, I.~Guyon, and V.~Vapnik, ``A training algorithm for optimal margin classifiers,'' in {\em Proceedings of the Fifth Annual Workshop on Computational Learning Theory}, (Pittsburgh), 1992.

\bibitem{murphy2012machine}
K.~P. Murphy, {\em Machine Learning: A Probabilistic Perspective}.
\newblock The MIT Press, 2012.

\bibitem{Pekalska09}
E.~P{\c e}kalska and B.~Haasdonk, ``Kernel discriminant analysis for positive definite and indefinite kernels,'' {\em IEEE Transactions on Pattern Analysis and Machine Intelligence}, vol.~31, no.~6, pp.~1017--1032, 2009.

\bibitem{scholkopf1998nonlinear}
B.~Schölkopf, A.~Smola, and K.~Müller, ``Nonlinear component analysis as a kernel eigen-value problem,'' {\em Neural Computation}, vol.~10, no.~5, pp.~1299--1319, 1998.

\bibitem{PANDEY2023126639}
A.~Pandey, H.~{De Meulemeester}, B.~{De Moor}, and J.~A. Suykens, ``Multi-view kernel pca for time series forecasting,'' {\em Neurocomputing}, vol.~554, p.~126639, 2023.

\bibitem{scholkopf1999input}
B.~Schölkopf, S.~Mika, C.~J.~C. Burges, P.~Knirsch, K.~R. Müller, G.~Raetsch, and A.~Smola, ``Input space vs. feature space in kernel-based methods,'' {\em IEEE Transactions on Neural Networks}, vol.~10, no.~5, pp.~1000--1017, 1999.

\bibitem{bach2002kernel}
F.~R. Bach and M.~I. Jordan, ``Kernel independent component analysis,'' {\em Journal of Machine Learning Research}, vol.~3, pp.~1--48, 2002.

\bibitem{schell2023nonlinear}
A.~Schell and H.~Oberhauser, ``Nonlinear independent component analysis for discrete-time and continuous-time signals,'' 2023.

\bibitem{shawe2004kernel}
J.~Shawe-Taylor and N.~Cristianini, {\em Kernel methods for pattern analysis}.
\newblock Cambridge university press, 2004.

\bibitem{langone2016kernel}
R.~Langone, R.~Mall, C.~Alzate, and J.~A. Suykens, ``Kernel spectral clustering and applications,'' {\em Unsupervised learning algorithms}, pp.~135--161, 2016.

\bibitem{HARVEY20123}
A.~Harvey and V.~Oryshchenko, ``Kernel density estimation for time series data,'' {\em International Journal of Forecasting}, vol.~28, no.~1, pp.~3--14, 2012.
\newblock Special Section 1: The Predictability of Financial Markets Special Section 2: Credit Risk Modelling and Forecasting.

\bibitem{lei2007astudy}
H.~Lei and B.~Sun, ``A study on the dynamic time warping in kernel machines,'' in {\em 2007 Third International IEEE Conference on Signal-Image Technologies and Internet-Based System}, pp.~839--845, 2007.

\bibitem{Badiane2022}
M.~Badiane and P.~Cunningham, ``An empirical evaluation of kernels for time series,'' {\em Artif Intell Rev}, vol.~55, pp.~1803--–1820, 2022.

\bibitem{smith1981identification}
T.~Smith and M.~Waterman, ``Identification of common molecular subsequences,'' {\em J. Mol. Biol.}, vol.~147, pp.~195--197, 1981.

\bibitem{LA04}
J.-P. Vert, H.~Saigo, and T.~Akutsu, ``Local alignment kernels for biologicalsequences,'' in {\em Kernel Methods in Computational Biology} (B.~Scholkopf, K.~Tsuda, and J.~Vert, eds.), p.~131–154, MIT Press, 2004.

\bibitem{haussler1999}
D.~Haussler, ``Convolution kernels on discrete structures,'' Technical Report UCSC-CRL-99-10, UC Santa Cruz, 1999.

\bibitem{haussler1999convolution}
D.~Haussler, ``Convolution kernels on discrete structures,'' Tech. Rep. UCSC-CRL-99-10, University of California in Santa Cruz, Computer Science Department, July 1999.

\bibitem{Jebara2004ProbabilityPK}
T.~Jebara, R.~Kondor, and A.~G. Howard, ``Probability product kernels,'' {\em J. Mach. Learn. Res.}, vol.~5, pp.~819--844, 2004.

\bibitem{NIPS2003_0abdc563}
P.~Moreno, P.~Ho, and N.~Vasconcelos, ``A kullback-leibler divergence based kernel for svm classification in multimedia applications,'' in {\em Advances in Neural Information Processing Systems} (S.~Thrun, L.~Saul, and B.~Sch\"{o}lkopf, eds.), vol.~16, MIT Press, 2003.

\bibitem{JMLR:v6:cuturi05a}
M.~Cuturi, K.~Fukumizu, and J.-P. Vert, ``Semigroup kernels on measures,'' {\em Journal of Machine Learning Research}, vol.~6, no.~40, pp.~1169--1198, 2005.

\bibitem{jaakkola1999fisher}
T.~Jaakkola, M.~Diekhans, and D.~Haussler, ``Using the fisher kernel method to detect remote protein homologies,'' in {\em Proceedings of the International Conference on Intelligent Systems for Molecular Biology}, August 1999.

\bibitem{jaakkola1999exploiting}
T.~Jaakkola and D.~Haussler, ``Exploiting generative models in discriminative classifiers,'' in {\em Advances in Neural Information Processing Systems}, pp.~487--493, 1999.

\bibitem{markov2023implementation}
V.~Markov, V.~Rastunkov, A.~Deshmukh, D.~Fry, and C.~Stefanski, ``Implementation and learning of quantum hidden markov models,'' 2023.

\bibitem{GM2020100285}
H.~GM, M.~K. Gourisaria, M.~Pandey, and S.~S. Rautaray, ``A comprehensive survey and analysis of generative models in machine learning,'' {\em Computer Science Review}, vol.~38, p.~100285, 2020.

\bibitem{jaeger2005learning}
H.~Jaeger, M.~Zhao, K.~Kretzschmar, T.~Oberstein, D.~Popovici, and A.~Kolling, ``Learning observable operator models via the es algorithm,'' {\em New directions in statistical signal processing: From systems to brains}, 2005.

\bibitem{carlyle_paz1971}
J.~Carlyle and A.~Paz, ``Realizations by stochastic finite automata,'' {\em Journal of Computer and System Sciences}, vol.~5, no.~1, pp.~26--40, 1971.

\bibitem{Dhanuka23}
R.~Dhanuka, J.~P. Singh, and A.~Tripathi, ``A comprehensive survey of deep learning techniques in protein function prediction,'' {\em IEEE/ACM Transactions on Computational Biology and Bioinformatics}, vol.~20, no.~3, pp.~2291--2301, 2023.

\bibitem{Jurgovsky2018SequenceCF}
J.~Jurgovsky, M.~Granitzer, K.~Ziegler, S.~Calabretto, P.-E. Portier, L.~He-Guelton, and O.~Caelen, ``Sequence classification for credit-card fraud detection,'' {\em Expert Syst. Appl.}, vol.~100, pp.~234--245, 2018.

\bibitem{Odaini23}
A.~A.~A. Odaini, F.~Zola, L.~Segurola-Gil, A.~Gil-Lertxundi, and C.~D’Andrea, ``Cybersecurity in public space: Leveraging cnn and lstm for proactive multivariate time series classification,'' in {\em 2023 IEEE International Conference on Big Data (BigData)}, pp.~4110--4118, 2023.

\bibitem{SharmaWeather21}
S.~Sharma, K.~K. Bhatt, R.~Chabra, and N.~Aneja, ``A comparative performance model of machine learning classifiers on time series prediction for weather forecasting,'' in {\em Advances in Information Communication Technology and Computing} (V.~Goar, M.~Kuri, R.~Kumar, and T.~Senjyu, eds.), (Singapore), pp.~577--587, Springer Nature Singapore, 2022.

\bibitem{LiangSM23}
M.~Liang, X.~Wang, and S.~Wu, ``Improving stock trend prediction through financial time series classification and temporal correlation analysis based on aligning change point,'' {\em Soft Comput}, vol.~27, p.~3655–3672, 2023.

\bibitem{Rosafalco20}
L.~Rosafalco, A.~Manzoni, and S.~Mariani, ``Fully convolutional networks for structural health monitoring through multivariate time series classification,'' {\em Adv. Model. and Simul. in Eng. Sci.}, vol.~7, p.~3655–3672, 2020.

\bibitem{Tsalikidis24}
N.~Tsalikidis, A.~Mystakidis, and C.~Tjortjis, ``Energy load forecasting: one-step ahead hybrid model utilizing ensembling,'' {\em Computing}, vol.~106, p.~241–273, 2024.

\bibitem{GOMES2023200268}
B.~Gomes, J.~Coelho, and H.~Aidos, ``A survey on traffic flow prediction and classification,'' {\em Intelligent Systems with Applications}, vol.~20, p.~200268, 2023.

\bibitem{monras2011hidden}
A.~Monras, A.~Beige, and K.~Wiesner, ``Hidden quantum markov models and non-adaptive read-out of many-body states,'' {\em Applied Mathematical and Computational Sciences}, vol.~3, no.~1, pp.~93--122, 2011.

\bibitem{nielsen_chuang_2010}
M.~A. Nielsen and I.~L. Chuang, {\em Quantum Computation and Quantum Information: 10th Anniversary Edition}.
\newblock Cambridge University Press, 2010.

\bibitem{Stinespring1955}
W.~F. Stinespring, ``Positive functions on {C}*-algebras,'' {\em Proceedings of the American Mathematical Society}, vol.~6, no.~2, p.~211–216, 1955.

\bibitem{scikit-learn}
F.~Pedregosa, G.~Varoquaux, A.~Gramfort, V.~Michel, B.~Thirion, O.~Grisel, M.~Blondel, P.~Prettenhofer, R.~Weiss, V.~Dubourg, J.~Vanderplas, A.~Passos, D.~Cournapeau, M.~Brucher, M.~Perrot, and E.~Duchesnay, ``Scikit-learn: Machine learning in {P}ython,'' {\em Journal of Machine Learning Research}, vol.~12, pp.~2825--2830, 2011.

\bibitem{Rebentrost2014Quantum}
P.~Rebentrost, M.~Mohseni, and S.~Lloyd, ``Quantum support vector machine for big data classification,'' {\em Phys. Rev. Lett.}, vol.~113, p.~130503, Sep 2014.

\bibitem{Chatterjee2017Generalized}
R.~Chatterjee and T.~Yu, ``Generalized coherent states, reproducing kernels, and quantum support vector machines,'' {\em Quantum Information and Computation}, vol.~17, Dec. 2017.

\bibitem{schuld2019quantum}
M.~Schuld and N.~Killoran, ``Quantum machine learning in feature hilbert spaces,'' {\em Physical review letters}, vol.~122, no.~4, p.~040504, 2019.

\bibitem{park2020practical}
J.-E. Park, B.~Quanz, S.~Wood, H.~Higgins, and R.~Harishankar, ``Practical application improvement to quantum svm: theory to practice,'' 2020.

\bibitem{rastunkov2022boosting}
V.~Rastunkov, J.-E. Park, A.~Mitra, B.~Quanz, S.~Wood, C.~Codella, H.~Higgins, and J.~Broz, ``Boosting method for automated feature space discovery in supervised quantum machine learning models,'' 2022.

\bibitem{schuld2021machine}
M.~Schuld and F.~Petruccione, {\em Machine learning with quantum computers}.
\newblock Springer, 2021.

\bibitem{cerezo2020variational}
M.~Cerezo, A.~Poremba, L.~Cincio, and P.~J. Coles, ``Variational quantum fidelity estimation,'' {\em Quantum}, vol.~4, p.~248, 2020.

\bibitem{wang2022quantum}
Q.~Wang, Z.~Zhang, K.~Chen, J.~Guan, W.~Fang, J.~Liu, and M.~Ying, ``Quantum algorithm for fidelity estimation,'' {\em IEEE Transactions on Information Theory}, vol.~69, no.~1, pp.~273--282, 2022.

\bibitem{chen2021variational}
R.~Chen, Z.~Song, X.~Zhao, and X.~Wang, ``Variational quantum algorithms for trace distance and fidelity estimation,'' {\em Quantum Science and Technology}, vol.~7, no.~1, p.~015019, 2021.

\bibitem{fuchs1996distinguishability}
C.~A. Fuchs, ``Distinguishability and accessible information in quantum theory,'' 1996.

\bibitem{jozsa1994fidelity}
R.~Jozsa, ``Fidelity for mixed quantum states,'' {\em Journal of modern optics}, vol.~41, no.~12, pp.~2315--2323, 1994.

\bibitem{liang2019quantum}
Y.-C. Liang, Y.-H. Yeh, P.~E. Mendon{\c{c}}a, R.~Y. Teh, M.~D. Reid, and P.~D. Drummond, ``Quantum fidelity measures for mixed states,'' {\em Reports on Progress in Physics}, vol.~82, no.~7, p.~076001, 2019.

\bibitem{Buhrman2001Quantum}
H.~Buhrman, R.~Cleve, J.~Watrous, and R.~de~Wolf, ``Quantum fingerprinting,'' {\em Phys. Rev. Lett.}, vol.~87, p.~167902, Sep 2001.

\bibitem{havlivcek2019supervised}
V.~Havl{\'\i}{\v{c}}ek, A.~D. C{\'o}rcoles, K.~Temme, A.~W. Harrow, A.~Kandala, J.~M. Chow, and J.~M. Gambetta, ``Supervised learning with quantum-enhanced feature spaces,'' {\em Nature}, vol.~567, no.~7747, pp.~209--212, 2019.

\bibitem{Huang2021Power}
H.-Y. Huang, M.~Broughton, M.~Mohseni, R.~Babbush, S.~Boixo, H.~Neven, and J.~R. McClean, ``Power of data in quantum machine learning,'' {\em Nature Communications}, vol.~12, no.~2631, 2021.

\bibitem{J__2022Quantum}
A.~J., A.~Adedoyin, J.~Ambrosiano, P.~Anisimov, W.~Casper, G.~Chennupati, C.~Coffrin, H.~Djidjev, D.~Gunter, S.~Karra, N.~Lemons, S.~Lin, A.~Malyzhenkov, D.~Mascarenas, S.~Mniszewski, B.~Nadiga, D.~O'malley, D.~Oyen, S.~Pakin, L.~Prasad, R.~Roberts, P.~Romero, N.~Santhi, N.~Sinitsyn, P.~J. Swart, J.~G. Wendelberger, B.~Yoon, R.~Zamora, W.~Zhu, S.~Eidenbenz, A.~Bärtschi, P.~J. Coles, M.~Vuffray, and A.~Y. Lokhov, ``Quantum algorithm~implementations for beginners,'' {\em {ACM} Transactions on Quantum Computing}, vol.~3, pp.~1--92, jul 2022.

\bibitem{das2019case}
P.~Das, S.~S. Tannu, P.~J. Nair, and M.~Qureshi, ``A case for multi-programming quantum computers,'' in {\em Proceedings of the 52nd Annual IEEE/ACM International Symposium on Microarchitecture}, pp.~291--303, 2019.

\bibitem{liu2021qucloud}
L.~Liu and X.~Dou, ``Qucloud: A new qubit mapping mechanism for multi-programming quantum computing in cloud environment,'' in {\em 2021 IEEE International symposium on high-performance computer architecture (HPCA)}, pp.~167--178, IEEE, 2021.

\bibitem{niu2023multi}
S.~Niu and A.~Todri-Sanial, ``Multi-programming mechanism on near-term quantum computing,'' in {\em Quantum Computing: Circuits, Systems, Automation and Applications}, pp.~19--54, Springer, 2023.

\end{thebibliography}

\newpage
\appendix

\section{Market Quantum Generative Models in Higher Dimensions}
\label{Appendix Higher Dim}
\begin{figure*}[ht]
        \includegraphics[width=1.\linewidth]{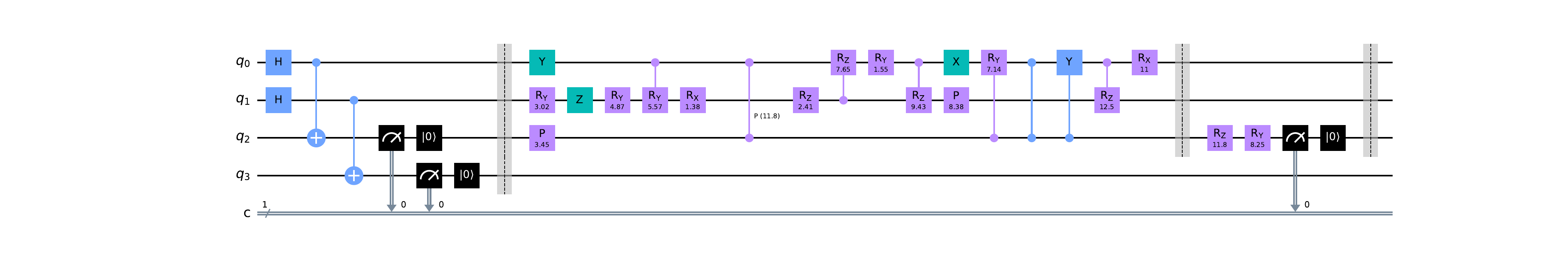}
         \caption{Market Movements QHMM: Two State Qubits; One Emission Qubit }
    \label{fig:example32fig1}
\end{figure*}

\begin{figure*}[ht]
        \includegraphics[width=1.\linewidth]{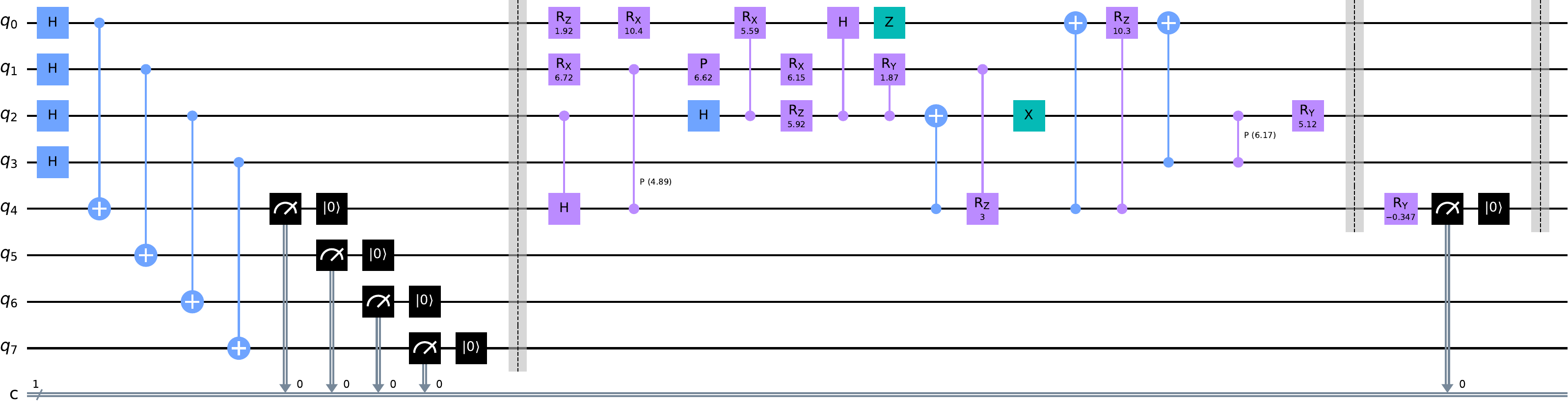}
         \caption{Market Movements QHMM: 4 State Qubits; One Emission Qubit}
    \label{fig:example33fig}
\end{figure*}

\section{Classification Tasks}
\label{Appendix Classifiaction tasks}
We empirically study binary classification problem for binary time series in the domain for market movement process (Example \ref{exm: market}). The generative model of the example domain is a QHMM (\ref{QHMM_D2}) parameterized in Hilbert spaces with dimensions 2, 4 and 8 
(Figures \ref{fig:example31fig}, \ref{fig:example32fig1}, \ref{fig:example33fig}).
 
\noindent
A structural classificataion task is defined by the following class-mapping function (\ref{clsmap_s}): 

\begin{equation}
p_s(\mathbf{y},c) = 
    \begin{cases*}
      1 & if $c=1 \land \sum_{i = 1}^{|\mathbf{y}|}y_i > \frac{|\mathbf{y}|}{2} $ \\
      0 & if $c=0 \land \sum_{i = 1}^{|\mathbf{y}|}y_i \leq \frac{|\mathbf{y}|}{2} $ 
    \end{cases*}
\end{equation}

The predictive classification task uses class-mapping function (\ref{clsmap_p}) defined for forward sequences with length $k=5$: 
$$p_p(\mathbf{y},k,c) = L[c_2],$$

where $c_2$ is the binary number corresponding to the prefix and the label of each prefix is defined by
$$L = '11110011001100100110001011000110' $$
The examples $\mathbf{y} \in \{0, 1\}^*$ are sampled with probability (\ref{eqn:seq_probability}):
\begin{equation*}
    P\bigl[ \mathbf{y} \vert \rho_0 \bigr] = \operatorname{tr}(T_{\mathbf{y}} \rho_0).
\end{equation*}

\newpage

\section {Kernel Induced Distances Across Classification Tasks - Trace Metric}
\label{Appendix Kernel Induced Distances}
\begin{figure}[ht]
    \begin{minipage}{.5\textwidth}
        \centering
        \includegraphics[trim = 0.0mm 0.0mm 0.0mm 0.0mm, clip,width=1.\linewidth]{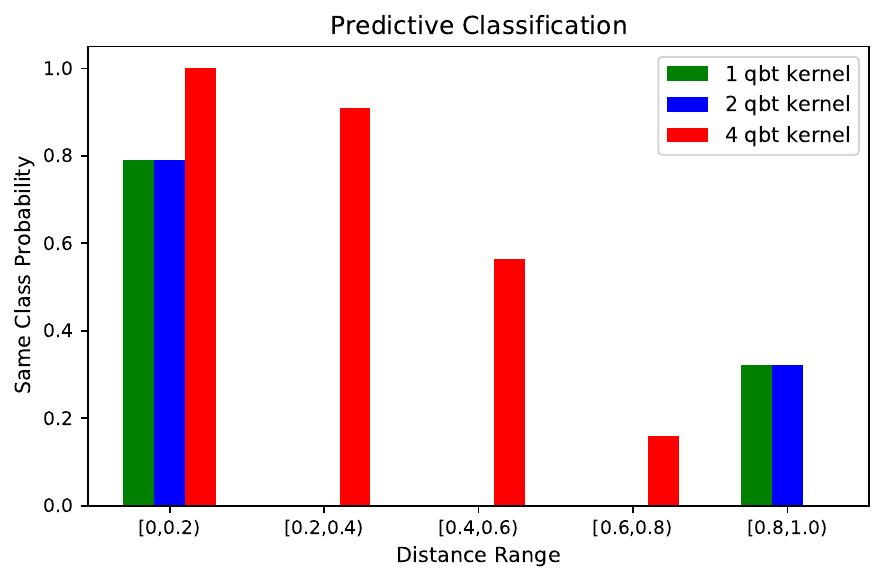}
        
        
    \end{minipage}
    \begin{minipage}{.5\textwidth}
        \centering
        
        \includegraphics[trim = 0.0mm 0.0mm 0.0mm 0.0mm, clip,width=1.\linewidth]{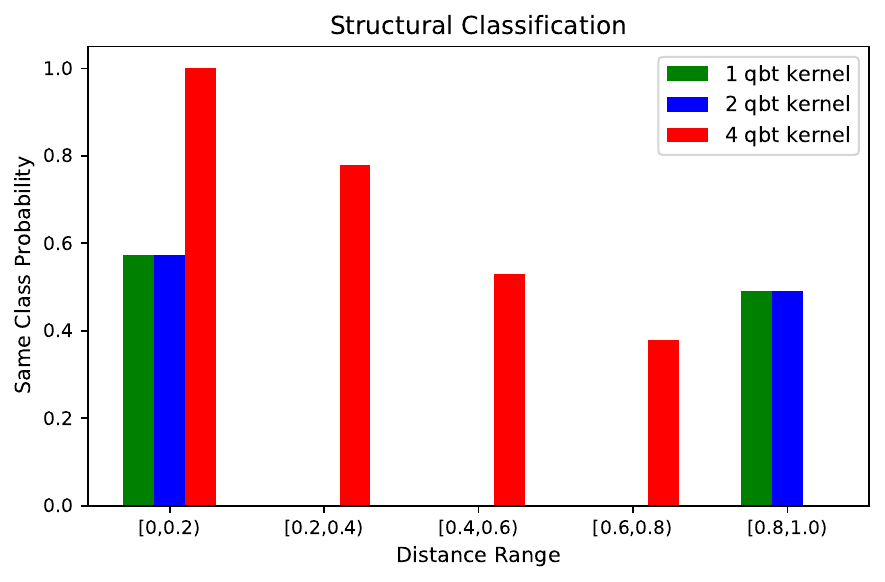}
        
    \end{minipage}
    \caption{Probability of Examples Belonging to the Same Class -Predictive Kernel, Trace Metric}
\label{fig:Trace_Struct_Kernel_Class}
\end{figure}

\begin{figure}[ht]
    \begin{minipage}{.5\textwidth}
        \centering
        \includegraphics[trim = 0.0mm 0.0mm 0.0mm 0.0mm, clip,width=1.\linewidth]{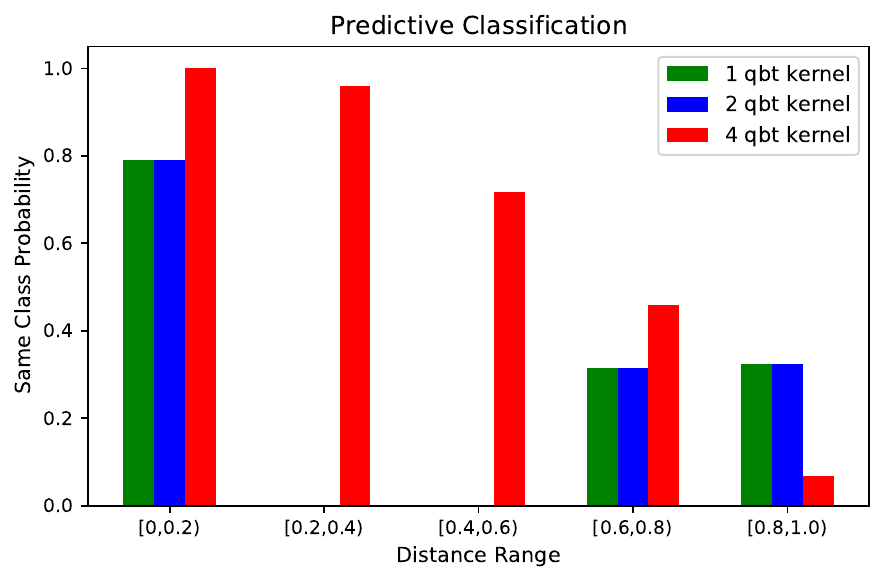}
        
        
    \end{minipage}
    \begin{minipage}{.5\textwidth}
        \centering
        
        \includegraphics[trim = 0.0mm 0.0mm 0.0mm 0.0mm, clip,width=1.\linewidth]{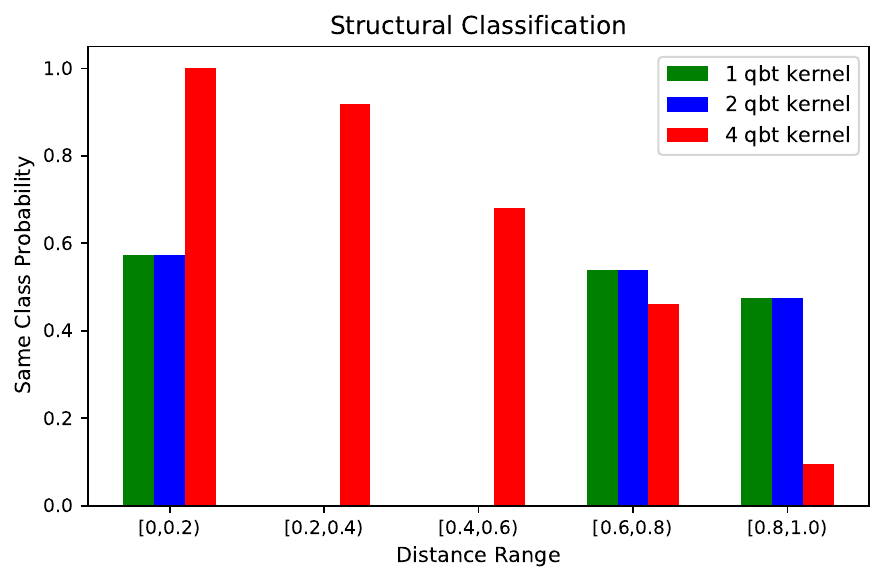}
        
    \end{minipage}
    \caption{Probability of Examples Belonging to the Same Class - Structural Kernel, Trace Metric}
\end{figure}

\newpage

\section{Kernels Performance - Distance Metric Bures}
\label{Appendix Kernels Performance - Distance Metric Bures}
    \begin{table*}[ht]
        \centering 
        \begin{tabular}{|c|c|c|c|c|c|}
            \hline
            Classifier & Kernel &  In Sample  & CI & Out Sample  & CI \\
            \hline
            RFS & N/A       & 0.985 & 0.984 - 0.987 & 0.824  &  0.819 - 0.829\\
            SVC & Classical & 0.962 & 0.959 - 0.965 & 0.903  &  0.896 - 0.909\\
            SVC & Quantum   & 0.998 & 0.997 - 0.998 & 0.947  &  0.944 - 0.949\\
            k-NN& Classical & 0.863 & 0.857 - 0.869 & 0.720  &  0.715 - 0.723 \\
            k-NN& Quantum   & 0.953 & 0.951 - 0.956 & 0.894  &  0.891 - 0.898 \\
            \hline
        \end{tabular}
            \caption{Kernels Performance by Recognizer Accuracy: Structural Task, Structural Kernel, Distance Metric- Bures, CI=95\%, Kernel Type RBF}
        \label{tab:structural_comparison_Bures}
    \end{table*}

    \begin{table*}[ht]
        \centering 
        \begin{tabular}{|c|c|c|c|c|c|}
            \hline
            Classifier & Kernel &  In Sample  & CI & Out Sample  & CI \\
            \hline
            RFS & N/A       & 0.998 & 0.996 - 0.999 & 0.968  &  0.966 - 0.971\\
            SVC & Classical & 0.951 & 0.947 - 0.955 & 0.916  &  0.910 - 0.922\\
            SVC & Quantum   & 0.999 & 0.998 - 0.999 & 0.977  &  0.975 - 0.979\\
            k-NN& Classical & 0.971 & 0.969 - 0.973 & 0.916  &  0.913 - 0.919 \\
            k-NN& Quantum   & 0.994 & 0.993 - 0.995 & 0.979  &  0.977 - 0.981 \\
            \hline
        \end{tabular}
            \caption{Kernels Performance by Recognizer Accuracy: Predictive Task, Predictive Kernel, Distance Metric- Bures, CI=95\%, Kernel Type RBF}
        \label{tab:predictive_comparison_Bures}
    \end{table*}

\end{document}